%% file: main.tex
\documentclass[lettersize,journal]{IEEEtran}

\usepackage[T1]{fontenc}
\usepackage{booktabs} 
\usepackage{graphicx}
\usepackage{float}
\usepackage{color}
\usepackage{array}
\usepackage{amssymb}
\usepackage{amsmath}
\usepackage{url}
\usepackage{subfig}
\usepackage{multirow}
\usepackage{hyperref}
\usepackage{pifont}
\usepackage{enumitem}
\usepackage{makecell} 
\usepackage[table]{xcolor}
\usepackage{colortbl}

\ifCLASSOPTIONcompsoc
  \usepackage[nocompress]{cite}
\else
  \usepackage{cite}
\fi
\ifCLASSINFOpdf
\else
\fi
\usepackage{makecell}

\newcommand{\reffig}[1]{{Fig.~\ref{#1}}}
\newcommand{\reftab}[1]{{Table~\ref{#1}}}

\begin{document}\sloppy

%
\title{Learn2Chat: Rethinking Dyadic Talking Heads via Interaction-Modulated Monologic Priors}

\author{
 	Zikai~Huang,
    Siyue~Chen,
 	Xuemiao~Xu, 
    Haoxin~Yang,
 	Cheng~Xu,
    Yihong~Lin, \\
 	and~Shengfeng He,~\IEEEmembership{Senior Member,~IEEE}
    \IEEEcompsocitemizethanks{
         \IEEEcompsocthanksitem \textup{Zikai Huang, Xuemiao Xu, Haoxin Yang, and Yihong Lin are with the School of Computer Science and Engineering, and Siyue Chen is with the School of Design, South China University of Technology, Guangdong, China.
         Xuemiao Xu is also with Guangdong Engineering Center for Large Model and GenAI Technology, and also with State Key Laboratory of Subtropical Building and Urban Science, Ministry of Education Key Laboratory of Big Data and Intelligent Robot. E-mail: zikaihuang0428@gmail.com; victoria38yue@gmail.com; harxis@outlook.com; amcsyihonglin@gmail.com; xuemx@scut.edu.cn.}
         \IEEEcompsocthanksitem \textup{Cheng Xu and Shengfeng He are with the School of Computing and Information Systems, Singapore Management University. Email: cschengxu@gmail.com; shengfenghe@smu.edu.sg.} 
    }
}
\markboth{IEEE Transactions on Visualization and Computer Graphics}%
{Huang \MakeLowercase{\textit{et al.}}: Think2Sing}
%




\makeatletter
\long\def\@IEEEtitleabstractindextextbox#1{\parbox{0.922\textwidth}{#1}}
\makeatother

\IEEEtitleabstractindextext{%

\input{sec/abstract}

}


\maketitle

\IEEEdisplaynontitleabstractindextext

%
\IEEEpeerreviewmaketitle

\input{fig/teaser}
\input{sec/intro}

\input{sec/related_work}
\input{sec/method}
\input{sec/experiment}
\input{sec/conclusion}


%





\ifCLASSOPTIONcaptionsoff
  \newpage
\fi

\bibliographystyle{IEEEtran}
\bibliography{main}



%



%

\input{sec/biography}





\end{document}


\sloppy

%
\title{Learn2Chat: Rethinking Dyadic Talking Heads via Interaction-Modulated Monologic Priors\\ - Supplementary Materials -}

\author{
 	Zikai~Huang,
    Siyue~Chen,
 	Xuemiao~Xu,
    Haoxin~Yang,
 	Cheng~Xu,
    Yihong~Lin, \\
 	and~Shengfeng He,~\IEEEmembership{Senior Member,~IEEE}
 }

\markboth{IEEE Transactions on Visualization and Computer Graphics}%
{Huang \MakeLowercase{\textit{et al.}}: Think2Sing}
%




\makeatletter
\long\def\@IEEEtitleabstractindextextbox#1{\parbox{0.922\textwidth}{#1}}
\makeatother


\maketitle

\IEEEdisplaynontitleabstractindextext

%
\IEEEpeerreviewmaketitle

\input{suppl/tab/listener}
\input{suppl/tab/generalization}
\input{suppl/intro}
\input{suppl/implementation_details}
\input{suppl/listener}
\input{suppl/generalization}


%
%
\bibliographystyle{splncs04}
\bibliography{suppl}

%% file: sec/abstract.tex
\begin{abstract}
Dyadic conversational motion generation is essential for realistic interactive digital humans. Existing approaches typically model conversational behaviors within unified dyadic generators. However, such holistic formulations tend to couple self-speech-driven motion with partner-responsive social feedback, leaving the interaction-specific component implicit and underutilizing the speech-motion correspondence already learned by pretrained monologic motion models.
We propose Learn2Chat, a unified framework that models dyadic motion as interaction modulation over pretrained monologic motion priors. This design separates intrinsic speech-driven motion from social interaction effects and enables more structured interaction modeling.
Specifically, we introduce a Monologic-Anchored Motion Factorization scheme that leverages the semantic motion manifold learned from monologic data to disentangle audio-driven motion dynamics from interaction-induced modulation, yielding clean interaction representations from dyadic sequences. 
On top of this representation space, a Cross-Attentive Interaction Latent Prediction module maps paired speech signals to interaction latents through cross-branch attention and interaction alignment. During inference, the predicted interaction latents modulate canonical monologic motion to generate coherent and synchronized dyadic behaviors in a data-efficient manner. 
Extensive experiments on the DualTalk benchmark demonstrate that Learn2Chat achieves state-of-the-art performance across both quantitative metrics and perceptual evaluations. Moreover, the framework is model-agnostic and seamlessly integrates with diverse pretrained monologic motion backbones, highlighting the effectiveness of prior reuse and interaction adaptation for scalable conversational motion generation.
More visual results are available on the \href{https://zikaihuangscut.github.io/Learn2Chat/}{\textcolor{blue}{project page}}.
\end{abstract}

%% file: fig/teaser.tex
\begin{figure*}[th]
    \centering
    \includegraphics[width=\linewidth,scale=1.00]{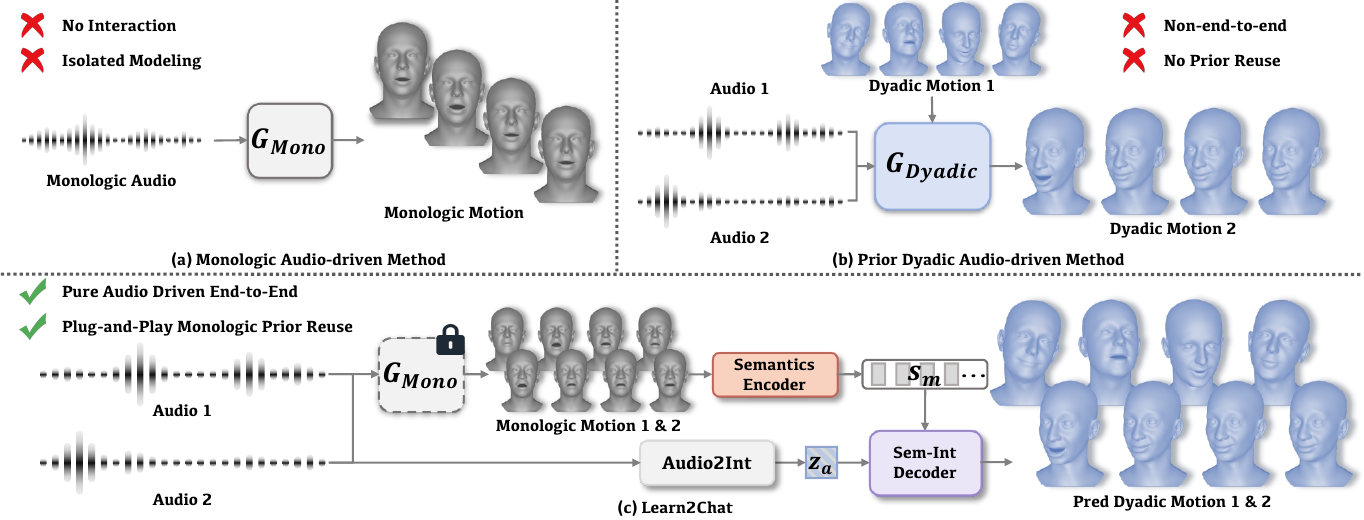}
    \caption{
        \textbf{Comparison of existing audio-driven head motion generation paradigms.}
         Monologic methods (a) ignore interaction dynamics, while existing dyadic approaches (b) either require auxiliary motion inputs or entangle speech and interaction signals through holistic learning. In contrast, Learn2Chat reformulates dyadic generation as interaction modulation over canonical monologic motion, enabling realistic and temporally coherent synthesis. The framework is plug-and-play and converts pretrained monologic backbones to the dyadic setting.}
    \label{teaser}
\end{figure*}

%% file: sec/intro.tex
\section{Introduction}
\IEEEPARstart{G}{enerating} conversational motions is essential for building lifelike interactive digital humans, with broad applications in telepresence~\cite{orts2016holoportation,lombardi2018deep}, virtual assistants~\cite{cassell2001embodied}, and autonomous social agents~\cite{hasenfuss2020reshaping}. Recent audio-driven motion generation methods have achieved remarkable fidelity in monologic settings~(\reffig{teaser}(a)), where facial movements and head motions are primarily driven by the speaker's own speech. Extending these advances to dyadic interaction, however, remains fundamentally challenging. Unlike monologic speech, conversations are inherently reciprocal: speakers continuously adjust their motions according to both their own utterances and their interlocutor's behaviors, producing temporally coordinated and socially adaptive motion patterns.

Existing dyadic motion generation methods can be broadly categorized into two paradigms. Early approaches rely on motion-conditioned generation, where conversational motion is synthesized by explicitly conditioning on the partner's motion signals~\cite{ng2022learning,tran2024dim,peng2025dualtalk}. Although effective, such methods require access to ground-truth partner motion and are therefore unsuitable for fully autonomous audio-driven applications. More recently, UniLS~\cite{chu2025unils} eliminates this dependency by generating dyadic motions from paired audio streams through end-to-end learning. Despite their differences in input modality and supervision, both paradigms share a common formulation: conversational behaviors are acquired by learning a dedicated dyadic generator from paired conversational data.

Meanwhile, monologic motion generation has witnessed rapid progress~\cite{fan2022faceformer,xing2023codetalker,peng2023selftalk,kim2025deeptalk,fan2024unitalker}. Trained on large speech-motion corpora, these models have learned effective self-speech-to-motion correspondence and can produce expressive facial and head motions from the speaker's own speech. This capability is directly relevant to dyadic generation, since a substantial part of conversational motion remains governed by the speaker's own utterance. Therefore, pretrained monologic models can provide more than a generic motion prior: they offer a self-speech-conditioned canonical motion reference, describing the motion already explained by the speaker's own speech.

This motivates a different view of dyadic motion generation. Conversational interaction does not replace the intrinsic speech-driven motion encoded in monologic models. Instead, it modulates how this motion should be expressed under different social contexts. In this view, dyadic motion can be regarded as canonical monologic motion enriched by interaction-induced adaptation, instead of an entirely new motion generation target. The central challenge is therefore to separate the motion already explained by the speaker's own speech from the interaction-specific modulation introduced by dyadic context. Without such a separation, intrinsic speech-driven motion and partner-responsive feedback remain entangled in a single dyadic generator, which may obscure the interaction-specific component and weaken the reuse of pretrained monologic priors.

To address this challenge, we propose \textbf{Learn2Chat}, a framework that extends pretrained monologic audio-driven models to dyadic conversations through interaction modulation. Specifically, we propose a Monologic-Anchored Motion Factorization scheme that distills interaction representations while preserving reusable speech-driven motion priors. To infer interaction directly from paired speech streams, we further introduce a Cross-Attentive Interaction Latent Predictor, which predicts interaction latents through cross-branch attention and interaction alignment. During inference, the predicted interaction latents modulate canonical monologic motion trajectories, enabling synchronized dyadic motion generation from paired audio alone.
Extensive experiments on the DualTalk benchmark demonstrate that Learn2Chat achieves state-of-the-art performance across both quantitative metrics and perceptual evaluations. More importantly, our framework is model-agnostic and seamlessly compatible with diverse pretrained monologic motion backbones, highlighting a new paradigm for scalable conversational motion generation through prior reuse and interaction adaptation.

In summary, our contributions are threefold:
\begin{itemize}[leftmargin=*, itemsep=0pt, topsep=0pt]
\item We propose a new formulation of dyadic motion generation as interaction modulation over pretrained monologic speech-motion priors, shifting the focus from unified speak-listen generation to modeling the interaction-induced adaptation beyond the motion already explained by the speaker's own speech. This encourages an explicit separation between intrinsic speech-driven motion and interaction-induced social feedback, thereby reducing signal conflation and learning ambiguity in holistic modeling.
\item We introduce Learn2Chat, a unified framework for dyadic motion generation. It employs a Monologic-Anchored Motion Factorization scheme to distill clean interaction representations from dyadic motion, together with a Cross-Attentive Interaction Latent Predictor that maps paired speech streams to interaction latents for modulating canonical monologic motions.
\item We demonstrate that Learn2Chat achieves state-of-the-art performance on the DualTalk dataset across quantitative and perceptual evaluations, while remaining model-agnostic and seamlessly compatible with existing monologic motion backbones.
\end{itemize}

%% file: sec/related_work.tex
\section{Related Work}

\noindent\textbf{Monologic Audio-driven 3D Head Generation.}
Audio-driven 3D talking head generation has advanced rapidly in recent years. Most existing approaches focus on monologic settings, where head motion is determined solely by the speaker's own audio~\cite{edwards2016jali,taylor2012dynamic,xu2013practical,li2017learning,fan2022faceformer,xing2023codetalker,peng2023selftalk,kim2025deeptalk,fan2024unitalker,cudeiro2019capture,chen2023diffusiontalker,peng2023emotalk,shen2024deitalk}. Early methods relied on explicit phoneme-to-viseme mappings constructed from handcrafted rules or graph-based matching~\cite{edwards2016jali,taylor2012dynamic,xu2013practical}. With the rise of deep learning, data-driven audio-to-motion modeling has become dominant. For instance, VOCA~\cite{cudeiro2019capture} regresses 3D facial motion using CNNs, while FaceFormer~\cite{fan2022faceformer} employs Transformers to capture long-range temporal dependencies. CodeTalker~\cite{xing2023codetalker} further introduces discrete latent representations via VQ-VAE to learn coupled audio–motion structures. More recently, diffusion-based approaches such as FaceDiffuser~\cite{stan2023facediffuser} improve motion realism and diversity. Despite their success, these methods focus on single-speaker scenarios and do not explicitly model conversational interaction. In contrast, our method leverages pretrained monologic motion priors while explicitly modeling interaction signals, enabling coherent dyadic generation.

\noindent\textbf{Dyadic Interaction Generation.}
Modeling dyadic interaction has recently gained increasing attention in conversational avatar research~\cite{ng2024audio,ng2022learning,tran2024dim,peng2025dualtalk,chu2025unils}. Early approaches often distinguish speaker and listener roles and train role-specific modules~\cite{ng2022learning,tran2024dim}, which typically require role annotations and limit generalization across conversational settings. More recent work seeks unified bidirectional modeling without explicit role supervision~\cite{peng2025dualtalk,chu2025unils}. For example, DualTalk~\cite{peng2025dualtalk} conditions generation on both participants' audio along with the partner's motion to capture conversational coordination. Related studies also explore contextual reaction modeling~\cite{ng2023can,wu2024vgg,zhou2022responsive}, though they often emphasize short-term responses and may struggle to maintain temporally coherent behaviors over longer interactions.
Recent work UniLS~\cite{chu2025unils} further advances this direction by removing dependence on partner motion and enabling fully audio-driven dyadic generation. Notably, UniLS adopts a two-stage training strategy, where the first stage learns an audio-independent motion prior from large-scale motion sequences, and the second stage aligns paired speech signals with this learned motion representation for dyadic synthesis. While this strategy improves optimization stability, the learned prior primarily captures general motion dynamics without explicitly encoding speech-conditioned semantic correspondence. As a result, interaction modeling and speech-driven motion generation are still learned jointly within a unified latent space during dyadic training, without an explicit factorization between intrinsic speech-motion mapping and interaction effects.

Instead of learning dyadic behavior as a unified generative model, we directly reuse a pretrained speech-conditioned motion model and keep its speech-motion mapping fixed. Dyadic interaction is then modeled as a separate module that modifies the generated motion conditioned on conversational context, rather than being entangled with motion generation itself. This leads to a clean separation between speech-driven motion generation and interaction modeling, enabling interaction to be learned independently as a lightweight add-on over a frozen monologic generator, rather than being absorbed into a jointly trained dyadic model.

\noindent\textbf{Content-Style Disentanglement in Motion Modeling.}
Disentangling content and style has been widely studied in motion generation to improve controllability and transferability. A common formulation separates motion into content factors that capture structural or semantic information and style factors that encode performer-specific or affective variations. In audio-driven head animation, prior work disentangles speech content from speaker identity~\cite{chai2022personalized,song2024talkingstyle,wang2023progressive,fu2024mimic,liang2025dgtalker,wang2025ptalker} or emotional expression~\cite{tan2024edtalk,peng2023emotalk,liu2024emoface,liu2024content}. In full-body motion synthesis, content typically represents action semantics while style models identity or emotional characteristics~\cite{aberman2020unpaired,song2024arbitrary,wu2024contrastive,zhong2024smoodi,kim2024most,wang2025difusion,petrovich2024multi}. Some work further integrates multimodal cues such as speech to generate stylized co-speech gestures and expressive body motion~\cite{ghorbani2023zeroeggs,ao2022rhythmic,liu2022disco,fu2024mambagesture}. Unlike these approaches, which focus on stylistic variation, we introduce a monologic-anchored factorization scheme that leverages pretrained monologic motion priors to separate speech-driven motion semantics from partner-conditioned interaction signals, enabling accurate modeling of conversational interaction for dyadic motion generation. 

%% file: sec/method.tex
\section{Method}
\subsection{Problem Formulation and Overview}
Let $(\mathbf{A}_1, \mathbf{A}_2) \in \mathbb{R}^{T \times D_a}$ denote the audio streams of two interlocutors in a dyadic conversation. Our goal is to synthesize temporally coherent and realistic dyadic 3D head-motion sequences $(\mathbf{x}^1_d, \mathbf{x}^2_d) \in \mathbb{R}^{T \times D_x}$.
We assume access to a pretrained monologic audio-driven head-motion generator $G_{\text{mono}}: \mathbf{A} \mapsto \mathbf{x}_m$ that produces speech-driven motion for a single speaker. Applying $G_{\text{mono}}$ independently to each audio stream yields speech-aligned base motions:
\begin{equation}
\mathbf{x}_m^1 = G_{\text{mono}}(\mathbf{A}_1), \qquad
\mathbf{x}_m^2 = G_{\text{mono}}(\mathbf{A}_2).
\end{equation}

Rather than learning a new end-to-end mapping from dual-speaker audio to dyadic motion from scratch, we reformulate dyadic generation as interaction modulation over these monologic bases. Specifically, a Monologic-Anchored Motion Factorization learning scheme learns a disentangled representation in which speech-driven semantics are decoupled from interaction patterns, while a Cross-Attentive Interaction Latent Prediction module predicts the interaction latent directly from $(\mathbf{A}_1, \mathbf{A}_2)$. As illustrated in~\reffig{overview}, we infer an interaction latent $\mathbf{z}$ from $(\mathbf{A}_1, \mathbf{A}_2)$ and generate the final motions via:
\begin{equation}
\hat{\mathbf{x}}^i_d = D\!\big(E_S(\mathbf{x}_m^i), \mathbf{z}^i, \mathbf{A}_i\big), \qquad i \in \{1, 2\},
\end{equation}
where $E_S(\cdot)$ encodes speech-driven motion semantics and $D(\cdot,\cdot,\cdot)$ is an audio-conditioned decoder that injects interaction patterns over canonical monologic priors via $\mathbf{z}$, modulating semantic dynamics to produce coherent interpersonal coordination while preserving speech alignment.
\input{fig/overview}

\subsection{Monologic-Anchored Motion Factorization}
We learn a monologic-anchored factorization that decomposes motion into (i) a shared semantic component capturing speech-driven dynamics
and (ii) an interaction component capturing partner-conditioned modulation. Concretely, both monologic and dyadic motion sequences are mapped into a
common semantic space by a shared encoder $E_S$. Interaction is modeled asymmetrically: dyadic motions use a dedicated interaction encoder $E_I$ to
infer partner-induced deviations, whereas monologic motions are associated with a learnable ``neutral-interaction'' prior. This design anchors
interaction learning to the robust semantic manifold provided by monologic priors and explicitly isolates partner-induced variation.

\noindent \textbf{Semantics Encoder.}
Given monologic motion $\mathbf{x}_m$ and dyadic motion $\mathbf{x}_d$, we construct a shared semantics encoder $E_S(\cdot)$ that maps both inputs into a common latent space:
\begin{equation}
\mathbf{s}_m = E_S(\mathbf{x}_m), 
\qquad 
\mathbf{s}_d = E_S(\mathbf{x}_d),
\end{equation}
where $\mathbf{s} \in \mathbb{R}^{T \times D_s}$ represents speech-driven motion semantics.
To reduce interaction-induced leakage at the input level, we first apply layer normalization to the motion sequence before feeding it into the temporal encoder.
This normalization mitigates scale and amplitude variations that are often associated with interaction intensity, thereby encouraging the encoder to focus on temporally structured motion semantics rather than magnitude-based stylistic cues.
The semantics encoder is implemented as a Transformer-based temporal model that captures long-range dependencies in motion trajectories.
Parameter sharing across monologic and dyadic inputs enforces domain invariance and prevents the encoder from specializing to a specific interaction regime.
Consequently, the learned semantics embedding captures motion patterns that are consistent across both settings, aligning with our interpretation of speech-driven base motion.

\noindent \textbf{Interaction Encoder.}
We model interaction as a stochastic latent variable. For dyadic motion, a Transformer encoder
$E_I(\cdot)$ predicts the parameters of a Gaussian distribution:
\begin{equation}
(\boldsymbol{\mu}_d, \boldsymbol{\sigma}_d) = E_I(\mathbf{x}_d),
\end{equation}
and samples the dyadic interaction latent via reparameterization:
\begin{equation}
\mathbf{z}_d = \boldsymbol{\mu}_d + \boldsymbol{\sigma}_d \odot \boldsymbol{\epsilon}, \qquad
\boldsymbol{\epsilon} \sim \mathcal{N}(\mathbf{0}, \mathbf{I}).
\end{equation}
A \textit{KL regularizer} encourages the posterior toward a standard normal prior:
\begin{equation}
\mathcal{L}_{\text{KL}} = D_{\text{KL}}\!\left(
\mathcal{N}\!\left(\boldsymbol{\mu}_d, (\boldsymbol{\sigma}_d)^2\right) \middle\|\,
\mathcal{N}\!\left(\mathbf{0}, \mathbf{I}\right)
\right).
\end{equation}
Monologic motion does not employ a data-driven interaction encoder. Instead, we introduce a learnable Gaussian prior $\mathbf{z}_m \sim \mathcal{N}(\mu_m, (\sigma_m)^2)$, where $\mu_m$ and
$\sigma_m$ are trainable parameters. This asymmetry reflects the modeling assumption that dyadic motion contains interaction-induced modulation, whereas monologic motion represents an
interaction-neutral baseline. By parameterizing the monologic interaction state as a compact learnable distribution, interaction variation is explicitly anchored to cross-domain discrepancy
rather than arbitrary intra-domain decomposition.

\noindent \textbf{Semantics-Interaction Decoder.}
Given a semantics embedding $\mathbf{s} \in \mathbb{R}^{T \times D_s}$, corresponding audio $\mathbf{A}$, and an interaction latent $\mathbf{z} \in \mathbb{R}^{D_z}$, the decoder reconstructs motion
$\hat{\mathbf{x}} = D(\mathbf{s}, \mathbf{z}, \mathbf{A})$, where $\hat{\mathbf{x}} \in \mathbb{R}^{T \times D_x}$.
The decoder comprises stacked Transformer blocks equipped with Adaptive Layer Normalization (AdaLN). In each block, the hidden state attends to pretrained Wav2Vec2 features
$\mathbf{a} \in \mathbb{R}^{T \times D_a}$ through cross-attention, while AdaLN conditions the normalization affine parameters on $\mathbf{z}$:
\begin{equation}
\mathbf{s}' = \mathbf{h} + \text{AdaLN}\!\Big(
\text{CA}\big(\mathbf{Q}=\mathbf{h},\, \mathbf{K}=\mathbf{a},\, \mathbf{V}=\mathbf{a}\big),\, \mathbf{z}
\Big).
\end{equation}
Conditioning normalization on $\mathbf{z}$ enables global modulation of motion dynamics without altering the attention operator, thereby preserving speech-driven structure while injecting
interaction-specific adjustments.

\noindent \textbf{Training Objective for Motion Factorization.}
To explicitly disentangle speech-driven semantics from interaction-induced modulation, we design a \emph{unified factorization loss} based on cross-domain recombination and cycle-consistent re-encoding. Specifically, we construct four recombinations with different semantics-interaction pairings to explicitly supervise the factorization of semantics and interaction:
\begin{equation}    
\begin{aligned}
\hat{\mathbf{x}}_{m2m} = D(\mathbf{s}_m, \mathbf{z}_m), \quad &\hat{\mathbf{x}}_{m2d} = D(\mathbf{s}_m, \mathbf{z}_d), \\
\hat{\mathbf{x}}_{d2m} = D(\mathbf{s}_d, \mathbf{z}_m), \quad &\hat{\mathbf{x}}_{d2d} = D(\mathbf{s}_d, \mathbf{z}_d).
\end{aligned}
\end{equation}

\noindent\emph{Reconstruction Loss.}
Domain-consistent reconstructions enforce fidelity to the original motion:
\begin{equation}
\mathcal{L}_{\text{rec}} = \| \hat{\mathbf{x}}_{m2m} - \mathbf{x}_m \|_2 + \| \hat{\mathbf{x}}_{d2d} - \mathbf{x}_d \|_2.
\end{equation}
This term ensures that semantics and interaction jointly preserve complete motion information.

\noindent\emph{Swap Loss.}
To explicitly attribute cross-domain discrepancy to the interaction latent, we supervise swapped combinations:
\begin{equation}
\mathcal{L}_{\text{swap}} = \| \hat{\mathbf{x}}_{m2d} - \mathbf{x}_d \|_2 + \| \hat{\mathbf{x}}_{d2m} - \mathbf{x}_m \|_2.
\end{equation}
This objective compels the interaction latent to encode interaction-dependent modulation sufficient to transform monologic motion into dyadic motion and vice versa.

\noindent\emph{Cycle Consistency.}
We further enforce factorization consistency by re-encoding synthesized motions.
Semantics consistency is imposed on all four outputs:
\begin{equation}
\mathcal{L}_{\text{sem-cyc}} = \sum_{i \in \{m2m,m2d,d2m,d2d\}} \| E_S(\hat{\mathbf{x}}_{i}) - \mathbf{s}^{\text{src}(i)} \|_2,
\end{equation}
where $\text{src}(i)$ denotes the original semantics source.
Since the interaction encoder is trained only on dyadic motion, interaction cycle consistency is enforced for dyadic-interaction outputs:
\begin{equation}
\mathcal{L}_{\text{int-cyc}} = \| E_I(\hat{\mathbf{x}}_{m2d}) - \mathbf{z}_d \|_2 + \| E_I(\hat{\mathbf{x}}_{d2d}) - \mathbf{z}_d \|_2.
\end{equation}
These constraints prevent information leakage between semantics and interaction representations.

\noindent \textit{Overall objective.} The final factorization loss is therefore defined as:
\begin{equation}
\mathcal{L}_{\text{fact}} =
\mathcal{L}_{\text{rec}}
+\lambda_{\text{swap}}\mathcal{L}_{\text{swap}}
+\lambda_{\text{sem}}\mathcal{L}_{\text{sem-cyc}}
+\lambda_{\text{int}}\mathcal{L}_{\text{int-cyc}}
+\lambda_{\text{KL}}\mathcal{L}_{\text{KL}}.
\end{equation}
This unified factorization objective yields a latent space in which dyadic interaction is represented by latents that are decoupled from speech-driven semantics.

\subsection{Cross-Attentive Dual-Audio Interaction Latent Prediction}
With the factorized motion representation learned above, we further learn a variational mapping from dual-speaker audio to the interaction manifold via a Cross-Attentive Dual-Audio Interaction Latent Prediction module. 

\noindent \textbf{Audio-to-Interaction Encoder.}
Let $\mathbf{a}_1,\mathbf{a}_2 \in \mathbb{R}^{T \times D_a}$ denote Wav2Vec2 features extracted from the two audio streams. In dyadic conversations, interaction is not a unary property attributable to either speaker alone; instead, it emerges from their mutual coordination and feedback. To capture such bidirectional dependency, we adopt a symmetric dual-stream Transformer with shared weights, in which intra-stream temporal modeling and inter-stream coupling are disentangled by design. Concretely, at each Transformer block, the two streams are updated in parallel as
\begin{equation}
\begin{aligned}
\mathbf{h}_1' &= \mathbf{h}_1 + \text{SA}(\mathbf{h}_1) + \text{CA}(\mathbf{Q}=\mathbf{h}_1,\mathbf{K}=\mathbf{h}_2,\mathbf{V}=\mathbf{h}_2), \\
\mathbf{h}_2' &= \mathbf{h}_2 + \text{SA}(\mathbf{h}_2) + \text{CA}(\mathbf{Q}=\mathbf{h}_2,\mathbf{K}=\mathbf{h}_1,\mathbf{V}=\mathbf{h}_1).
\end{aligned}
\end{equation}
Here, $\text{SA}$ captures the temporal evolution within each speaker (e.g., prosodic dynamics and long-range acoustic structure), while $\text{CA}$ explicitly conditions each stream on its interlocutor, encoding reciprocal coupling. This architectural separation prevents holistic entanglement of within-speaker dynamics and cross-speaker interaction cues, and encourages the model to represent interaction as relational modulation rather than independent temporal patterns. We prepend a learnable \texttt{[CLS]} token to each stream to aggregate global conversational context. After $L$ layers, the \texttt{[CLS]} embeddings are projected to Gaussian parameters, from which the audio-conditioned interaction latent $\mathbf{z}_a$ is sampled via reparameterization to match the stochastic interaction formulation learned in the motion space.

\noindent \textbf{Training Objective for Interaction Latent Prediction.}
During training, we freeze the motion factorization model (i.e., $E_S$, $E_I$, and $D$) learned in the previous step. Given a monologic semantics embedding $\mathbf{s}_m$ and the predicted audio interaction latent $\mathbf{z}_a$, we generate dyadic motion as
\begin{equation}
\hat{\mathbf{x}}_{a\to d} = D(\mathbf{s}_m,\mathbf{z}_a).
\end{equation}

\noindent\textit{Motion Reconstruction Loss.}
We supervise the generated motion against the ground-truth dyadic motion:
\begin{equation}
\mathcal{L}_{\text{motion}} = \|\hat{\mathbf{x}}_{a\to d} - \mathbf{x}_d\|_2.
\end{equation}
This term ensures that the predicted interaction latent induces motion modulation consistent with the dyadic interaction dynamics observed in data.

\noindent\textit{Cross-modal Interaction Alignment Loss.}
To avoid degenerate solutions where the audio encoder explains dyadic motion solely through reconstruction while drifting away from the intended interaction manifold, we explicitly align audio-derived interaction embeddings with motion-derived dyadic interaction embeddings using a contrastive InfoNCE loss:
\begin{equation}
\mathcal{L}_{\text{InfoNCE}}
= -\log
\frac{\exp(\text{sim}(\mathbf{z}_a^{i}, \mathbf{z}_d^{i})/\tau)}
{\sum_{j}\exp(\text{sim}(\mathbf{z}_a^{i}, \mathbf{z}_d^{j})/\tau)},
\end{equation}
where $\text{sim}(\cdot,\cdot)$ denotes cosine similarity and $\tau$ is a temperature parameter. This contrastive objective aligns the audio-conditioned interaction distribution with the interaction manifold distilled from dyadic motions, enforcing latent-level consistency such that interaction inferred from audio corresponds to the same modulation variable used to transform canonical monologic motions.

\noindent\textit{Overall objective.} The final objective of interaction latent prediction is as:
\begin{equation}
\mathcal{L}_{\text{pred}} = \mathcal{L}_{\text{motion}} + \lambda_{\text{InfoNCE}} \mathcal{L}_{\text{InfoNCE}}.
\end{equation}
During inference, the inferred interaction latent $\mathbf{z}_a$ is combined with monologic semantics $\mathbf{s}_m$ to generate dyadic motion via the decoder, enabling fully audio-driven dyadic generation while remaining plug-and-play compatible with different monologic models.

%% file: fig/overview.tex
\begin{figure*}[t]
    \centering
    \includegraphics[width=\linewidth,scale=1.00]{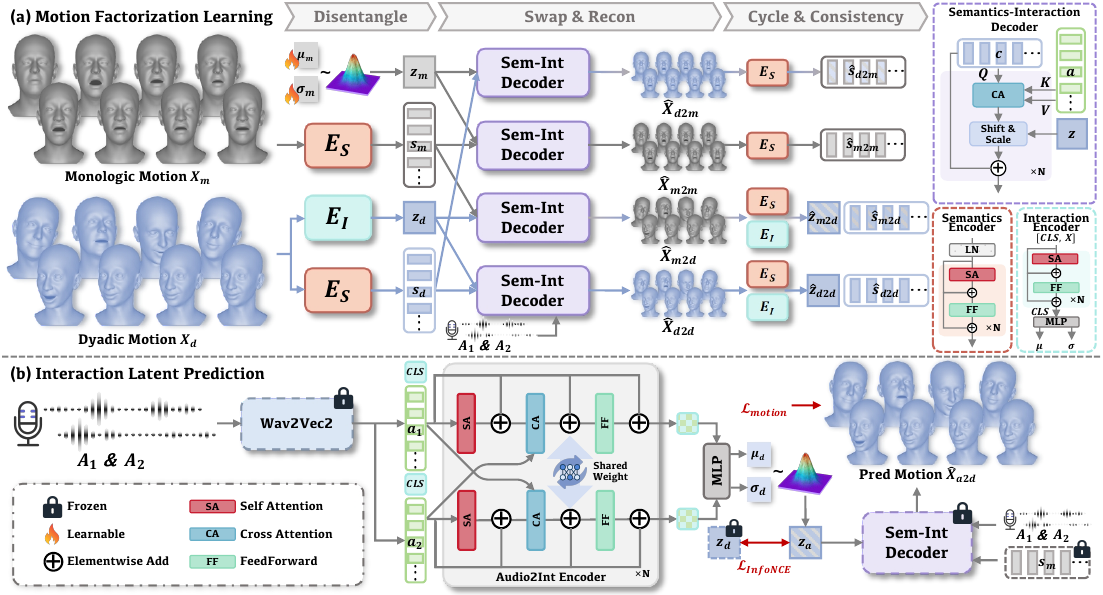}
    \caption{
        \textbf{Overview of Learn2Chat.}
        Learn2Chat reformulates dyadic head motion generation as interaction modulation of pretrained monologic motion priors. 
        (a) The Monologic-Anchored Motion Factorization Learning explicitly disentangles speech-driven semantics from interaction.
        (b) The Cross-Attentive Interaction Latent Prediction module infers interaction latent from dual-speaker audio via a dual-stream encoder, enabling plug-and-play dyadic synthesis without retraining the monologic generator.
    }
    \label{overview}
\end{figure*}

%% file: sec/experiment.tex
\section{Experiment}
\subsection{Settings}
\noindent \textbf{Dataset.} We evaluate Learn2Chat on the DualTalk dataset~\cite{peng2025dualtalk}, a benchmark tailored for audio-driven dyadic head motion generation.
DualTalk contains over 50 hours of multi-round conversations spanning more than 1{,}000 identities, with isolated audio streams and corresponding FLAME parameters~\cite{li2017learning}.
It covers diverse interaction scenarios where speaker--listener roles naturally alternate over time.
We follow the official data split and report results on both the standard test set and the out-of-distribution (OOD) set to assess generalization.

\noindent \textbf{Metrics.} Following prior protocols~\cite{ng2022learning,tran2024dim,peng2025dualtalk}, we adopt a comprehensive suite of quantitative metrics that measure motion fidelity, prediction accuracy, diversity, and interpersonal temporal coherence.
Specifically, motion fidelity is measured by Fr\'echet Distance (FD) and Paired Fr\'echet Distance (P-FD).
Prediction accuracy is evaluated by Mean Squared Error (MSE) between generated and ground-truth motion parameters.
Motion diversity is quantified by SI for Diversity (SID).
Inter-speaker temporal coordination is assessed using the residual Pearson Correlation Coefficient (rPCC).
For more implementation details, we kindly refer readers to the supplementary material.

\input{tab/comparison_test}
\input{tab/comparison_ood}
\input{tab/params}
\subsection{Comparison with State-of-the-art Methods}
We compare Learn2Chat with representative state-of-the-art approaches for audio-driven head motion generation in conversational settings.
The baselines can be grouped into two categories:

\textit{Monologic audio-driven methods.}
The first group consists of monologic generators that synthesize head motion from a single audio stream, including FaceFormer~\cite{fan2022faceformer}, CodeTalker~\cite{xing2023codetalker}, SelfTalk~\cite{peng2023selftalk}, DEEPTalk~\cite{kim2025deeptalk}, and UniTalker~\cite{fan2024unitalker}.
These methods are designed for single-speaker settings and do not explicitly model interpersonal interaction.
For a fair comparison, we retrain these models on DualTalk using the same data split and training protocol.
At inference, each participant is synthesized independently from its corresponding audio stream.
For methods whose results are directly reported under the DualTalk evaluation setup (FaceFormer~\cite{fan2022faceformer}, CodeTalker~\cite{xing2023codetalker}, and SelfTalk~\cite{peng2023selftalk}), we additionally include the published numbers when directly comparable.

\textit{Dyadic interaction methods.}
The second group consists of dyadic interaction models, including L2L~\cite{ng2022learning}, DIM~\cite{tran2024dim}, DualTalk~\cite{peng2025dualtalk}, and the recent audio-only framework UniLS~\cite{chu2025unils}. These methods differ in their inference protocols. L2L, DIM, and DualTalk require the partner's motion as an auxiliary condition during inference, whereas UniLS directly generates both participants' motions from paired audio streams without requiring motion inputs.

To ensure a fair comparison under the audio-only deployment setting considered in this work, we adapt only the motion-conditioned methods by replacing the required partner motion with monologic base motions generated by a pretrained monologic model $G_{\text{mono}}$. Specifically, we employ two representative monologic generators, DEEPTalk~\cite{kim2025deeptalk} and UniTalker~\cite{fan2024unitalker}, using their publicly released pretrained checkpoints to generate the motion conditions. This protocol removes the dependency on unavailable ground-truth partner motions while preserving the original inference pipeline of these methods as closely as possible. In contrast, UniLS is evaluated using its original audio-only inference protocol without modification. All baselines are retrained on the DualTalk dataset using the same training split, preprocessing pipeline, and evaluation protocol for a consistent comparison.

\reftab{comparison_test} and \reftab{comparison_ood} summarize the quantitative results on the DualTalk test and OOD splits.
Across both evaluation settings and under different monologic backbones, Learn2Chat consistently achieves the best overall performance.
Notably, the gains are reflected in motion fidelity, interaction coherence, and temporal coordination, and remain stable under distribution shifts, demonstrating the robustness of our interaction-modulation formulation.

\subsubsection{Comparison with Monologic Methods}
Monologic baselines generate each participant independently and therefore lack explicit modeling of interpersonal feedback.
Consequently, they underperform Learn2Chat across all metrics and motion components.
In particular, FD and P-FD are consistently higher, indicating a clear distributional gap between independently generated motions and real dyadic behaviors.
Inter-speaker coordination is especially weak, as evidenced by substantially worse rPCC.
While MSE is relatively less degraded than distributional and coordination metrics, it remains consistently inferior to Learn2Chat, suggesting that independent synthesis does not reliably improve even frame-level parameter accuracy.
Moreover, although some methods yield non-trivial SID, the absence of structured interaction modeling limits this variability from manifesting as meaningful conversational diversity.
Overall, these results confirm that monologic generation alone is insufficient for realistic dyadic head motion synthesis.

\subsubsection{Comparison with Dyadic Interaction Methods}
We compare with representative dyadic interaction methods, which can be broadly categorized into motion-conditioned approaches and end-to-end audio-driven approaches. 
Motion-conditioned methods, including L2L~\cite{ng2022learning}, DIM~\cite{tran2024dim}, and DualTalk~\cite{peng2025dualtalk}, generate conversational motion by explicitly conditioning on the partner's motion signal. When adapted to the audio-only setting by replacing ground-truth partner motion with monologic base motions, these methods exhibit consistent degradation across FD, P-FD, MSE, and rPCC, indicating a strong reliance on high-fidelity motion supervision and limited robustness under realistic deployment conditions. 
Although SID can appear relatively higher in some cases, it is often accompanied by noticeable temporal jitter and inconsistent motion trajectories, suggesting that such gains are primarily driven by increased stochastic variation rather than improved interaction modeling. 
In contrast, UniLS~\cite{chu2025unils} removes the dependency on partner motion and simultaneously learns dyadic motions from paired audio streams.
However, this joint learning strategy entangles motion generation and interaction modeling within a unified latent space and requires large-scale dyadic training data, which can make optimization less stable under limited-data scenarios. In our experiments, this is also reflected in performance degradation compared to strong motion-conditioned baselines such as DualTalk~\cite{peng2025dualtalk}, suggesting that coupling motion and interaction learning may not be optimal when training data is constrained.
Our approach benefits from explicitly reusing pretrained monologic speech-motion priors, providing a more stable and transferable motion foundation, which leads to improved motion fidelity and more reliable cross-speaker coordination, especially under OOD evaluation. 
Overall, motion-conditioned methods struggle with missing motion supervision, while end-to-end dyadic models are data-hungry and tend to entangle motion and interaction learning. 
In comparison, Learn2Chat achieves more robust performance by leveraging prior reuse and decoupling interaction modeling from motion generation through structured adaptation over a pretrained monologic motion space.
We further compare the number of trainable parameters among different dyadic methods, as summarized in \reftab{params}.
Although our model contains slightly more parameters than L2L~\cite{ng2022learning}, it achieves substantially better performance.
In contrast to DIM~\cite{tran2024dim}, DualTalk~\cite{peng2025dualtalk} and UniLS~\cite{chu2025unils}, our approach is significantly more parameter-efficient, requiring less than 42\%, 8\% and 11\% of their trainable parameters, respectively.
These results demonstrate a favorable balance between effectiveness and efficiency.

\input{fig/qualitative_results}
\subsubsection{Qualitative Comparison}
\reffig{qualitative_results} provides a qualitative comparison.
Compared with existing methods, Learn2Chat produces more natural motions with improved interaction coherence, better reflecting realistic conversational dynamics and cross-speaker temporal coordination.
More visual results are available on the \href{https://zikaihuangscut.github.io/Learn2Chat/}{\textcolor{blue}{project page}}.

\input{fig/user_study}
\input{tab/ablation}
\input{tab/ablation_recon}
To assess the perceptual quality of the generated head motions, we conduct a user study comparing our method with several baseline approaches.
The evaluation focuses on three aspects of conversational animation: lip synchronization, interaction naturalness, and overall motion quality.
We randomly select a set of 20 conversational audio clips from the Test / OOD split of the DualTalk dataset~\cite{peng2025dualtalk} and generate the corresponding head motion sequences using our method and the baseline models.
All sequences are rendered with the same 3D head model, lighting, and camera configuration to ensure consistent visual conditions. Each clip contains the full interaction between two speakers and lasts around 20 $\sim$ 30 seconds to ensure sufficient context for evaluation.
For each trial, participants watch video clips generated by different methods for the same conversation segment.
The presentation order is randomized to avoid ordering bias.
Participants independently rate each clip according to three criteria.
Lip synchronization measures how well the head motion aligns with the rhythm and articulation of speech.
Interaction naturalness evaluates whether the two speakers exhibit realistic conversational coordination.
Overall quality reflects the general realism and visual plausibility of the animation.
All ratings are collected using a five point Likert scale, where 1 indicates poor quality and 5 indicates excellent quality.
A total of 25 participants with normal or corrected to normal vision took part in the study.
The results in~\reffig{user_study} show that our method consistently receives higher scores across all evaluation criteria, with the most noticeable improvement in interaction naturalness and overall quality.
Compared with monologic methods~(DEEPTalk~\cite{kim2025deeptalk}, UniTalker~\cite{fan2024unitalker}), our method shows significant improvements in interaction naturalness while preserving decent lip synchronization, which can be attributed to our explicit modeling of interaction dynamics and the disentangled representation learning that captures the underlying semantics of speech and interaction.
For some dyadic methods like L2L~\cite{ng2022learning} and DIM~\cite{tran2024dim}, without extra input of interlocutor's motion as an anchor, the generated head motions struggle to capture interaction dynamics and shows significant degradation across all criteria.
DualTalk~\cite{peng2025dualtalk} performs better than other non-end-to-end dyadic baselines, but still falls short of our method  in both motion fidelity and coordination quality.
UniLS~\cite{chu2025unils} provides a strong end-to-end audio-driven baseline.
However, with limited dyadic training data, its joint modeling of motion generation and interaction within a unified latent space tends to entangle speech-driven motion semantics with interaction cues, which may limit explicit modeling of interaction structure and lead to less precise coordination compared to our approach.
This outcome indicates that the proposed framework produces more realistic and coordinated conversational head motions than the compared methods.

\subsection{Ablation Study}
In this section, we conduct in-depth ablations to validate the contribution of our key design choices.

\noindent \textbf{Effectiveness of the Interaction-Modulated Framework.}
We examine the necessity of interaction modulation by removing the motion factorization stage~(w/o Motion Factorization) and learning the audio-to-interaction mapping directly.
As shown in \reftab{ablation}, this variant consistently degrades motion fidelity and inter-speaker coordination.
This indicates that, without an explicit factorized interaction space anchored by monologic priors, the model is prone to conflating speech-driven dynamics with partner-conditioned modulation, leading to unstable and less coherent conversational behaviors.
In contrast, learning an interaction latent that modulates canonical monologic motions provides a structured pathway to encode interpersonal cues while preserving speech alignment.

\noindent \textbf{Effectiveness of Monologic-Anchored Motion Factorization Learning Scheme.}
We next ablate the objectives used to learn the proposed monologic-anchored factorization, isolating the contribution of each loss term.
Since this factorization defines the disentangled semantics-interaction space used for interaction modulation, its reconstruction quality directly reflects how well interaction is separated from speech-driven motion.
As shown in \reftab{ablation_recon}, removing $\mathcal{L}_{rec}$ causes the most severe degradation, confirming that domain-consistent reconstruction is essential to preserve motion information while stabilizing the decomposition.
Excluding $\mathcal{L}_{swap}$ mainly worsens FD/P-FD, validating its role in enforcing cross-domain recombination and attributing domain discrepancy to the interaction latent.
Removing $\mathcal{L}_{sem\mbox{-}cyc}$ or $\mathcal{L}_{int\mbox{-}cyc}$ yields moderate drops, indicating that cycle-consistency constraints help suppress information leakage and improve robustness.
Finally, ablating $\mathcal{L}_{KL}$ slightly degrades FD/P-FD and rPCC while inflating SID, suggesting that distribution regularization encourages a compact interaction manifold and reduces noise-induced variability.

\noindent \textbf{Effectiveness of Cross-Attentive Interaction Latent Prediction Module.}
We further validate the cross-attentive interaction latent prediction module by replacing the proposed dual-stream encoder with a single-stream variant~(w Single-Stream Encoder) that concatenates the two speakers' audio features and feeds them into a shared encoder.
As shown in \reftab{ablation}, this replacement leads to a pronounced performance drop, particularly on coordination-related metrics.
This supports our design choice that dual-stream cross-attention is critical for capturing bidirectional acoustic dependencies and predicting interaction as relational coupling, rather than conflating within-speaker temporal dynamics with cross-speaker cues in a single entangled stream.

%% file: tab/comparison_test.tex
\begin{table*}[t]
\caption{
    \textbf{Comparison of our method with existing approaches on the DualTalk dataset test set.}
    \textbf{Bold} indicates best result. $\downarrow$ means lower is better, $\uparrow$ means higher is better.
}
\label{comparison_test}
\renewcommand{\arraystretch}{1.}
\setlength{\tabcolsep}{3pt}
\resizebox{\textwidth}{!}{\begin{tabular}{@{}c|ccc|ccc|ccc|ccc|ccc@{}}
\bottomrule
\multirow{3}{*}{\textbf{Methods}}           & \multicolumn{3}{c|}{\textbf{FD $\downarrow$}}                                                                                                          & \multicolumn{3}{c|}{\textbf{P-FD $\downarrow$}}                                                                                                        & \multicolumn{3}{c|}{\textbf{MSE $\downarrow$}}                                                                                                                                                       & \multicolumn{3}{c|}{\textbf{SID $\uparrow$}} & \multicolumn{3}{c}{\textbf{rPCC $\downarrow$}}                                                                                                                                                                  \\ \cline{2-16} 
                                            & exp               & \begin{tabular}[c]{@{}c@{}}jaw\\ $\times 10^{-3}$ \end{tabular} & \begin{tabular}[c]{@{}c@{}}pose\\ $\times 10^{-2}$ \end{tabular} & exp               & \begin{tabular}[c]{@{}c@{}}jaw\\ $\times 10^{-3}$ \end{tabular} & \begin{tabular}[c]{@{}c@{}}pose\\ $\times 10^{-2}$ \end{tabular} & \begin{tabular}[c]{@{}c@{}}exp\\ $\times 10^{-1}$ \end{tabular} & \begin{tabular}[c]{@{}c@{}}jaw\\ $\times 10^{-3}$ \end{tabular} & \begin{tabular}[c]{@{}c@{}}pose\\ $\times 10^{-2}$ \end{tabular} & exp              & jaw              & pose             & \begin{tabular}[c]{@{}c@{}}exp\\ $\times 10^{-2}$ \end{tabular} & \begin{tabular}[c]{@{}c@{}}jaw\\ $\times 10^{-1}$ \end{tabular} & \begin{tabular}[c]{@{}c@{}}pose\\ $\times 10^{-1}$ \end{tabular} \\ \hline
FaceFormer~\cite{fan2022faceformer}         & 34.90             & 5.40                                                            & 8.00                                                             & 34.90             & 5.40                                                            & 8.00                                                             & 6.97                                                            & 1.80                                                            & 2.67                                                             & 0.54             & 0.36             & 0.50             & 13.05                                                            & 2.41                                                            & 5.27                                                             \\
CodeTalker~\cite{xing2023codetalker}        & 48.57             & 6.89                                                            & 10.74                                                            & 48.57             & 6.89                                                            & 10.74                                                            & 9.71                                                            & 2.29                                                            & 3.58                                                             & 0.00             & 0.00             & 0.00             & 11.06                                                            & 2.33                                                            & 5.11                                                             \\
SelfTalk~\cite{peng2023selftalk}            & 35.77             & 5.49                                                            & 8.14                                                             & 35.77             & 5.49                                                            & 8.14                                                             & 7.15                                                            & 1.83                                                            & 2.71                                                             & 2.49             & 1.30             & 1.28             & 12.25                                                            & 2.39                                                            & 4.70                                                             \\
DEEPTalk~\cite{kim2025deeptalk}             & 28.24             & 4.04                                                            & 7.70                                                             & 28.85             & 4.09                                                            & 7.78                                                             & 6.32                                                            & 1.59                                                            & 2.73                                                             & 2.58             & 1.42             & 1.27             & 11.57                                                            & 2.26                                                            & 4.36                                                             \\
UniTalker~\cite{fan2024unitalker}           & 28.04             & 3.71                                                            & 7.75                                                             & 28.67             & 3.76                                                            & 7.83                                                             & 6.34                                                            & 1.48                                                            & 2.72                                                             & 2.35             & 1.89             & 1.24             & 11.44                                                            & 2.18                                                            & 4.33                                                             \\ \hline
L2L~\cite{ng2022learning}(DEEPTalk)         & 26.10             & 42.95                                                           & \textbf{0.99}                                                    & 27.03             & 43.02                                                           & \textbf{1.93}                                                    & 17.14                                                           & 24.25                                                           & 5.83                                                             & 4.06             & \underline{2.54} & \textbf{3.11}    & 12.78                                                            & 6.05                                                            & 3.17                                                             \\
L2L~\cite{ng2022learning}(UniTalker)        & 23.63             & 36.35                                                           & \underline{1.66}                                                 & 24.61             & 36.41                                                           & \underline{2.65}                                                 & 15.78                                                           & 21.80                                                           & 6.30                                                             & 4.17             & \textbf{2.63}    & \underline{3.05} & 11.29                                                            & 5.32                                                            & 3.23                                                             \\
DIM~\cite{tran2024dim}(DEEPTalk)            & 35.31             & 5.00                                                            & 7.90                                                             & 36.03             & 5.00                                                            & 7.90                                                             & 7.84                                                            & 2.00                                                            & 2.70                                                             & \textbf{4.59}    & 1.92             & 1.02             & 12.50                                                            & 2.59                                                            & 5.40                                                             \\
DIM~\cite{tran2024dim}(UniTalker)           & 38.16             & 4.98                                                            & 8.32                                                             & 38.79             & 5.04                                                            & 8.37                                                             & 8.13                                                            & 1.79                                                            & 2.84                                                             & \underline{4.50} & 1.93             & 1.12             & 13.20                                                            & 2.74                                                            & 5.96                                                             \\
DualTalk~\cite{peng2025dualtalk}(DEEPTalk)  & 20.32             & 3.24                                                            & 6.42                                                             & 20.64             & 3.28                                                            & 6.47                                                             & 4.64                                                            & \underline{1.30}                                                & 2.25                                                             & 3.01             & 1.90             & 1.22             & 7.86                                                             & 1.93                                                            & 3.87                                                             \\
DualTalk~\cite{peng2025dualtalk}(UniTalker) & 21.16             & 3.25                                                            & 6.37                                                             & 21.46             & 3.29                                                            & 6.42                                                             & 4.78                                                            & 1.35                                                            & 2.24                                                             & 3.02             & 2.02             & 1.22             & 7.65                                                             & 1.89                                                            & 3.94                                                             \\
UniLS~\cite{chu2025unils}                   & 25.48             & 3.14                                                            & 7.96                                                             & 26.45             & 3.20                                                            & 8.20                                                             & 6.22                                                            & 1.48                                                            & 3.38                                                             & 2.79             & 2.61             & 1.08             & 7.33                                                             & 2.26                                                            & 4.41                                                             \\ \hline
\rowcolor{gray!20}Ours(DEEPTalk)            & \textbf{14.95}    & \textbf{1.83}                                                   & 5.78                                                             & \textbf{15.96}    & \textbf{1.93}                                                   & 5.97                                                             & \textbf{4.60}                                                   & \textbf{1.16}                                                   & \underline{2.19}                                                 & 3.12             & 2.25             & 1.38             & \textbf{6.59}                                                    & \textbf{1.34}                                                   & \underline{2.58}                                                 \\
\rowcolor{gray!20}Ours(UniTalker)           & \underline{15.58} & \underline{1.84}                                                & 5.46                                                             & \underline{16.58} & \underline{1.95}                                                & 5.62                                                             & \underline{4.62}                                                & \textbf{1.16}                                                   & \textbf{2.18}                                                    & 3.10             & 2.25             & 1.49             & \underline{6.99}                                                 & \underline{1.38}                                                & \textbf{2.49}                                                    \\ \toprule
\end{tabular}}                
\end{table*}

%% file: tab/comparison_ood.tex
\begin{table*}[t]
\caption{
    \textbf{Comparison of our method with existing approaches on the DualTalk dataset OOD set.}
    \textbf{Bold} indicates best result. $\downarrow$ means lower is better, $\uparrow$ means higher is better.
}
\label{comparison_ood}
\renewcommand{\arraystretch}{1.}
\setlength{\tabcolsep}{3pt}
\resizebox{\textwidth}{!}{\begin{tabular}{@{}c|ccc|ccc|ccc|ccc|ccc@{}}
\bottomrule
\multirow{3}{*}{\textbf{Methods}}           & \multicolumn{3}{c|}{\textbf{FD $\downarrow$}}                                                                                                          & \multicolumn{3}{c|}{\textbf{P-FD $\downarrow$}}                                                                                                        & \multicolumn{3}{c|}{\textbf{MSE $\downarrow$}}                                                                                                                                                       & \multicolumn{3}{c|}{\textbf{SID $\uparrow$}}                   & \multicolumn{3}{c}{\textbf{rPCC $\downarrow$}}                                                                                                                                                        \\ \cline{2-16} 
                                            & exp               & \begin{tabular}[c]{@{}c@{}}jaw\\ $\times 10^{-3}$ \end{tabular} & \begin{tabular}[c]{@{}c@{}}pose\\ $\times 10^{-2}$ \end{tabular} & exp               & \begin{tabular}[c]{@{}c@{}}jaw\\ $\times 10^{-3}$ \end{tabular} & \begin{tabular}[c]{@{}c@{}}pose\\ $\times 10^{-2}$ \end{tabular} & \begin{tabular}[c]{@{}c@{}}exp\\ $\times 10^{-1}$ \end{tabular} & \begin{tabular}[c]{@{}c@{}}jaw\\ $\times 10^{-3}$ \end{tabular} & \begin{tabular}[c]{@{}c@{}}pose\\ $\times 10^{-2}$ \end{tabular} & exp              & jaw              & pose                     & \begin{tabular}[c]{@{}c@{}}exp\\ $\times 10^{-2}$ \end{tabular} & \begin{tabular}[c]{@{}c@{}}jaw\\ $\times 10^{-1}$ \end{tabular} & \begin{tabular}[c]{@{}c@{}}pose\\ $\times 10^{-1}$ \end{tabular} \\ \hline
FaceFormer~\cite{fan2022faceformer}         & 35.92             & 5.39                                                            & 8.60                                                             & 35.92             & 5.39                                                            & 8.60                                                             & 7.18                                                            & 1.80                                                            & 2.87                                                             & 0.54              & 0.40             & 0.51                    & 11.71                                                            & 2.16                                                            & 5.73                                                             \\
CodeTalker~\cite{xing2023codetalker}        & 50.05             & 6.95                                                            & 11.66                                                            & 50.05             & 6.95                                                            & 11.66                                                            & 10.01                                                           & 2.32                                                            & 3.88                                                             & 0.00              & 0.00             & 0.00                    & 10.24                                                            & 2.18                                                            & 5.76                                                             \\
SelfTalk~\cite{peng2023selftalk}            & 36.23             & 5.36                                                            & 8.89                                                             & 36.23             & 5.36                                                            & 8.89                                                             & 7.24                                                            & 1.79                                                            & 2.96                                                             & 2.61              & 1.36             & 1.08                    & 11.26                                                            & 2.13                                                            & 5.67                                                             \\
DEEPTalk~\cite{kim2025deeptalk}             & 29.87             & 3.63                                                            & 8.04                                                             & 30.47             & 3.68                                                            & 8.12                                                             & 6.70                                                            & 1.50                                                            & 2.88                                                             & 2.36              & 1.96             & 1.15                    & 10.17                                                            & 1.89                                                            & 5.61                                                             \\
UniTalker~\cite{fan2024unitalker}           & 30.08             & 3.98                                                            & 8.18                                                             & 30.69             & 4.04                                                            & 8.26                                                             & 6.73                                                            & 1.58                                                            & 2.89                                                             & 2.35              & 1.76             & 1.12                    & 10.53                                                            & 1.98                                                            & 5.58                                                             \\ \hline
L2L~\cite{ng2022learning}(DEEPTalk)         & 27.13             & 42.76                                                           & \textbf{1.30}                                                    & 29.53             & 42.81                                                           & \textbf{2.16}                                                    & 17.40                                                           & 24.14                                                           & 6.06                                                             & 3.91              & \underline{2.61} & \textbf{3.12}           & 15.50                                                            & 6.00                                                            & 3.52                                                             \\
L2L~\cite{ng2022learning}(UniTalker)        & 24.63             & 36.26                                                           & \underline{2.14}                                                 & 26.99             & 36.32                                                           & \underline{3.02}                                                 & 15.99                                                           & 21.69                                                           & 6.50                                                             & \underline{4.09}  & \textbf{2.68}    & \underline{3.05}        & 13.90                                                            & 5.36                                                            & 3.56                                                             \\
DIM~\cite{tran2024dim}(DEEPTalk)            & 36.59             & 5.00                                                            & 8.20                                                             & 37.29             & 5.00                                                            & 8.30                                                             & 8.12                                                            & 2.00                                                            & 2.80                                                             & \textbf{4.27}     & 2.00             & 1.45                    & 13.10                                                            & 2.50                                                            & 5.80                                                             \\
DIM~\cite{tran2024dim}(UniTalker)           & 41.37             & 4.86                                                            & 8.76                                                             & 42.03             & 4.91                                                            & 8.81                                                             & 8.79                                                            & 1.75                                                            & 2.98                                                             & 3.59              & 1.93             & 0.77                    & 13.30                                                            & 2.54                                                            & 6.32                                                             \\
DualTalk~\cite{peng2025dualtalk}(DEEPTalk)  & 27.58             & 3.69                                                            & 7.91                                                             & 27.92             & 3.73                                                            & 7.95                                                             & 6.16                                                            & 1.47                                                            & 2.74                                                             & 2.77              & 1.88             & 1.56                    & 9.04                                                             & 1.85                                                            & 5.12                                                             \\
DualTalk~\cite{peng2025dualtalk}(UniTalker) & 28.17             & 3.93                                                            & 7.90                                                             & 31.11             & 3.97                                                            & 7.94                                                             & 6.78                                                            & 1.60                                                            & 2.75                                                             & 2.65              & 1.91             & 1.34                    & 9.22                                                             & 1.82                                                            & 4.95                                                             \\ 
UniLS~\cite{chu2025unils}                   & 26.18             & 2.92                                                            & 7.79                                                             & 27.11             & 2.98                                                            & 8.02                                                             & 6.40                                                            & 1.42                                                            & 3.38                                                             & 2.84             & 2.62             & 1.05             & 7.68                                                             & 1.92                                                            & 5.03                                                             \\ \hline
\rowcolor{gray!20}Ours(DEEPTalk)            & \textbf{22.34}    & \underline{2.51}                                                & 6.00                                                             & \textbf{23.41}    & \underline{2.61}                                                & 6.20                                                             & \underline{6.24}                                                & \underline{1.42}                                                & \underline{2.42}                                                 & 2.79              & 2.09             & 1.36                    & \textbf{7.72}                                                    & \underline{1.46}                                                & \underline{2.31}                                                 \\
\rowcolor{gray!20}Ours(UniTalker)           & \underline{22.55} & \textbf{2.50}                                                   & 5.62                                                             & \underline{23.60} & \textbf{2.59}                                                   & 5.79                                                             & \textbf{6.16}                                                   & \textbf{1.40}                                                   & \textbf{2.22}                                                    & 2.79              & 2.07             & 1.48                    & \underline{7.88}                                                 & \textbf{1.46}                                                   & \textbf{2.22}                                                    \\ \toprule
\end{tabular}}
\end{table*}

%% file: tab/params.tex
\begin{table*}[t]
\centering
\caption{
\textbf{Number of trainable parameters with dyadic methods.}
Our model strides for a favorable balance between performance and efficiency.
}
\label{params}
\renewcommand{\arraystretch}{1}
\begin{tabular}{@{}cccccc@{}}
\toprule
\textbf{Methods}     & Ours   & L2L~\cite{ng2022learning}   & DIM~\cite{tran2024dim}    & DualTalk~\cite{peng2025dualtalk} & UniLS~\cite{chu2025unils} \\ \midrule
\textbf{\# Params/M} & 48.13  & 18.47                       & 114.93                    & 638.82                           & 421.3                     \\ \bottomrule
\end{tabular}
\end{table*}

%% file: fig/qualitative_results.tex
\begin{figure*}[tb]
    \centering
    \includegraphics[width=\linewidth,scale=1.00]{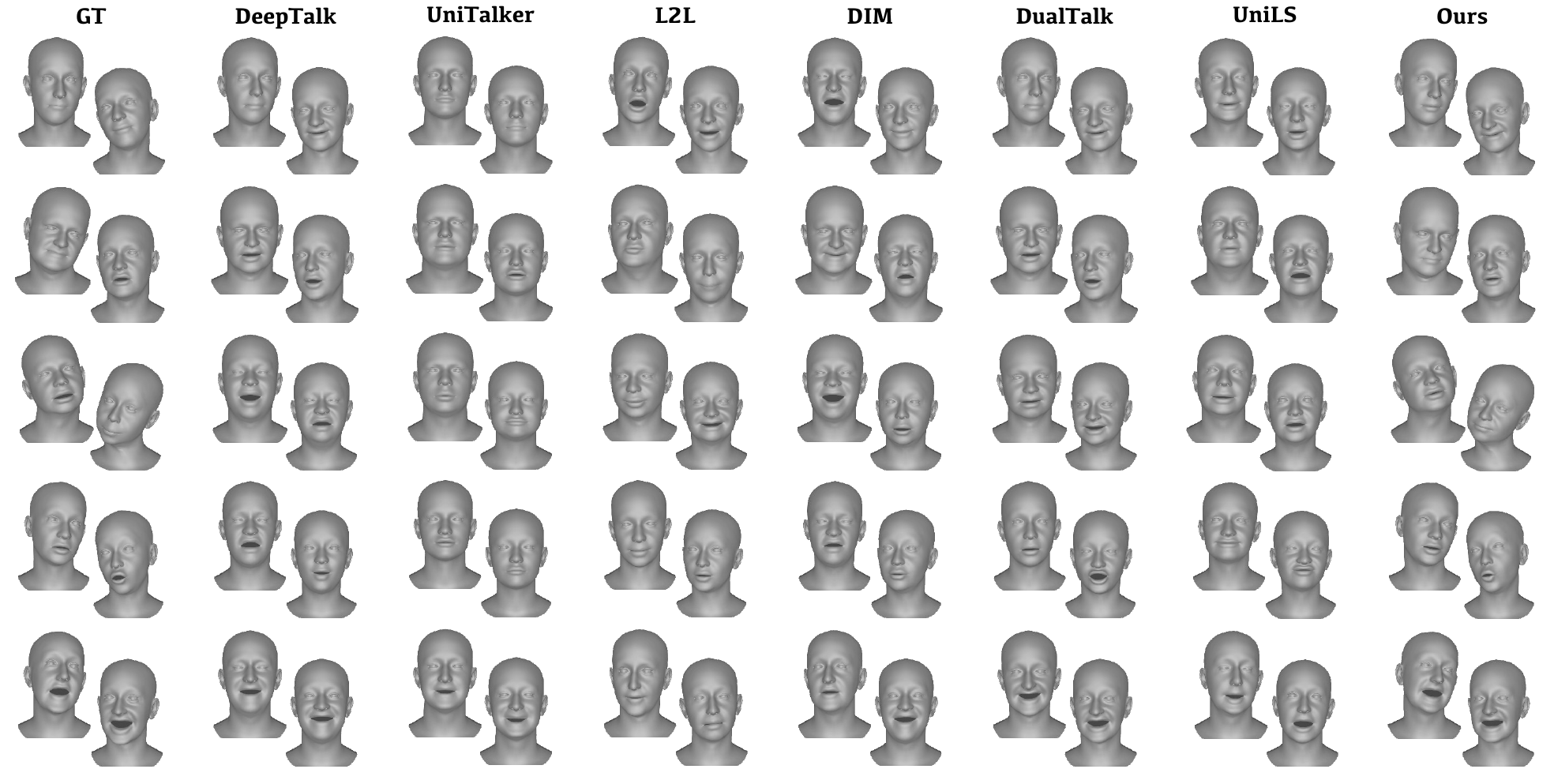}
    \caption{
        \textbf{Qualitative comparison of generated dyadic head motion sequences across different methods.}
        Our approach produces more natural and interaction-coherent behaviors that better align with real conversational dynamics.
    }
    \label{qualitative_results}
\end{figure*}

%% file: fig/user_study.tex
\begin{figure}[tb]
    \centering
    \centering
    \includegraphics[width=.45\textwidth]{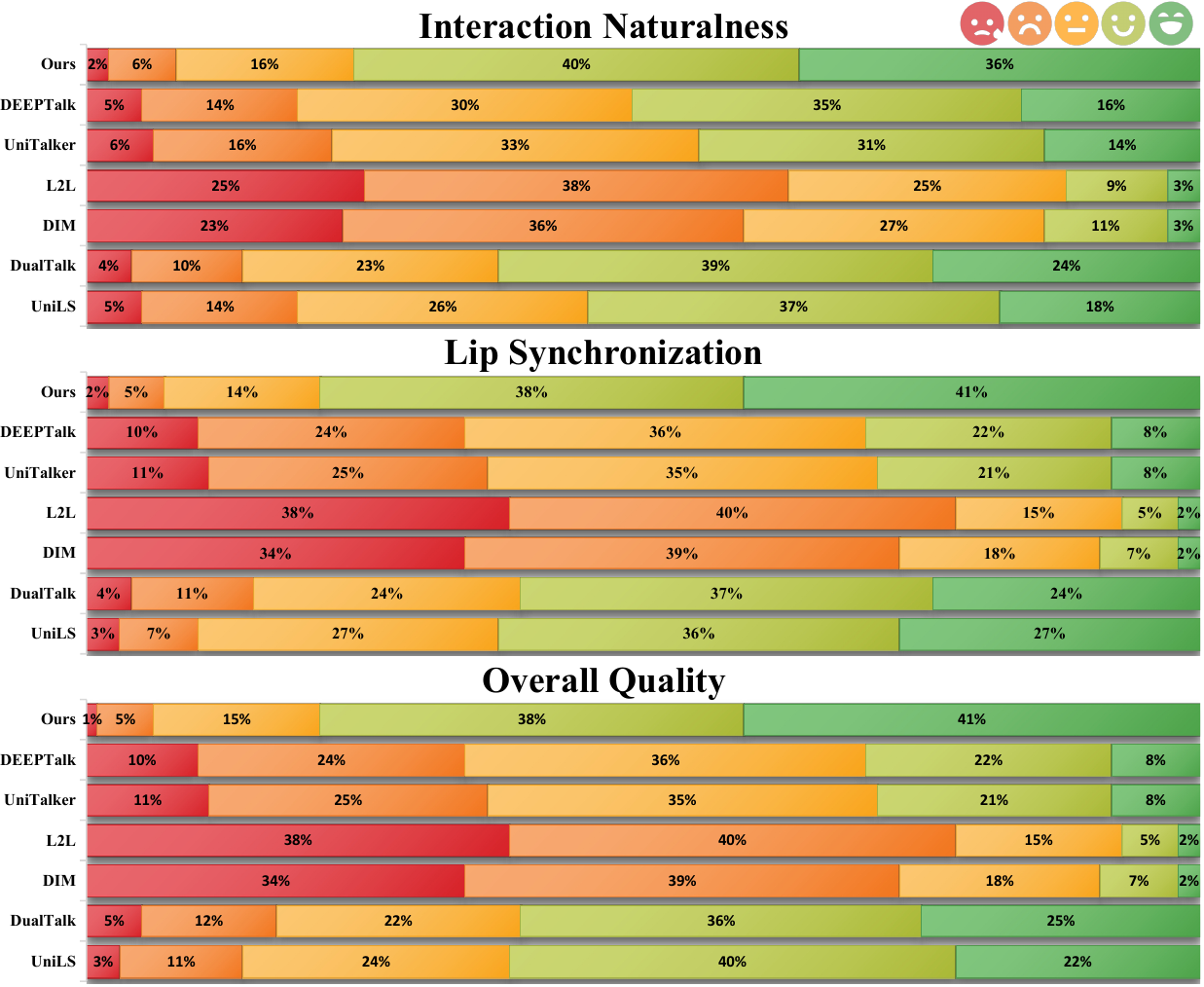}
    \caption{
        \textbf{User study on lip synchronization, interaction naturalness, and overall quality.}
        Our method consistently outperforms the compared methods across all evaluation criteria.
    }
    \label{user_study}
\end{figure}

%% file: tab/ablation.tex
\begin{table*}[th]
\caption{
    \textbf{Ablation study on motion factorization and interaction latent prediction design.}
    The results demonstrate the effectiveness of the proposed monologic-anchored motion factorization and cross-attentive Interaction latent prediction module.
}
\label{ablation}
\resizebox{\textwidth}{!}{\begin{tabular}{@{}cc|ccc|ccc|ccc|ccc|ccc@{}}
\bottomrule
\multicolumn{2}{c|}{}                                                                       & \multicolumn{3}{c|}{\textbf{FD $\downarrow$}}                                                                           & \multicolumn{3}{c|}{\textbf{P-FD $\downarrow$}}                                                                       & \multicolumn{3}{c|}{\textbf{MSE $\downarrow$}}                                                                        & \multicolumn{3}{c|}{\textbf{SID $\uparrow$}}                                                                          & \multicolumn{3}{c}{\textbf{rPCC $\downarrow$}}                                                                  \\ \cline{3-17} 
\multicolumn{2}{c|}{\multirow{-2}{*}{\textbf{Methods}}}                                     & \textbf{exp}                           & \textbf{jaw}                          & \textbf{pose}                         & \textbf{exp}                           & \textbf{jaw}                          & \textbf{pose}                         & \textbf{exp}                          & \textbf{jaw}                          & \textbf{pose}                         & \textbf{exp}                          & \textbf{jaw}                          & \textbf{pose}                         & \textbf{exp}                          & \textbf{jaw}                          & \textbf{pose}                         \\ \hline
\multicolumn{1}{c|}{}                                & \cellcolor[HTML]{F2F2F2}Ours         & \cellcolor[HTML]{F2F2F2}\textbf{14.95} & \cellcolor[HTML]{F2F2F2}\textbf{1.83} & \cellcolor[HTML]{F2F2F2}\textbf{5.78} & \cellcolor[HTML]{F2F2F2}\textbf{15.96} & \cellcolor[HTML]{F2F2F2}\textbf{1.93} & \cellcolor[HTML]{F2F2F2}\textbf{5.97} & \cellcolor[HTML]{F2F2F2}\textbf{4.60} & \cellcolor[HTML]{F2F2F2}\textbf{1.16} & \cellcolor[HTML]{F2F2F2}\textbf{2.19} & \cellcolor[HTML]{F2F2F2}\textbf{3.12} & \cellcolor[HTML]{F2F2F2}\textbf{2.25} & \cellcolor[HTML]{F2F2F2}\textbf{1.38} & \cellcolor[HTML]{F2F2F2}\textbf{6.59} & \cellcolor[HTML]{F2F2F2}\textbf{1.34} & \cellcolor[HTML]{F2F2F2}\textbf{2.58} \\
\multicolumn{1}{c|}{}                                & w/o Motion Factorization             & 24.53                                  & 3.20                                  & 6.47                                  & 25.27                                  & 3.27                                  & 6.55                                  & 6.01                                  & 1.51                                  & 2.63                                  & 2.63                                  & 2.56                                  & 1.12                                  & 10.28                                 & 1.99                                  & 4.00                                  \\ 
\multicolumn{1}{c|}{\multirow{-3}{*}{\textbf{Test}}} & w Single-Stream Encoder              & 19.42                                  & 2.11                                  & 6.70                                  & 18.51                                  & 2.23                                  & 6.79                                  & 5.34                                  & 1.31                                  & 2.37                                  & 2.74                                  & 2.56                                  & 1.01                                  & 8.07                                  & 1.69                                  & 3.14                                  \\ \hline \hline
\multicolumn{1}{c|}{}                                & \cellcolor[HTML]{F2F2F2}Ours         & \cellcolor[HTML]{F2F2F2}\textbf{22.34} & \cellcolor[HTML]{F2F2F2}\textbf{2.51} & \cellcolor[HTML]{F2F2F2}\textbf{6.00} & \cellcolor[HTML]{F2F2F2}\textbf{23.41} & \cellcolor[HTML]{F2F2F2}\textbf{2.61} & \cellcolor[HTML]{F2F2F2}\textbf{6.20} & \cellcolor[HTML]{F2F2F2}\textbf{6.24} & \cellcolor[HTML]{F2F2F2}\textbf{0.60} & \cellcolor[HTML]{F2F2F2}\textbf{2.42} & \cellcolor[HTML]{F2F2F2}\textbf{2.79} & \cellcolor[HTML]{F2F2F2}\textbf{2.09} & \cellcolor[HTML]{F2F2F2}\textbf{1.36} & \cellcolor[HTML]{F2F2F2}\textbf{7.72} & \cellcolor[HTML]{F2F2F2}\textbf{1.46} & \cellcolor[HTML]{F2F2F2}\textbf{2.31} \\
\multicolumn{1}{c|}{}                                & w/o Motion Factorization             & 26.57                                  & 3.12                                  & 8.15                                  & 27.30                                  & 3.19                                  & 8.03                                  & 6.98                                  & 1.61                                  & 2.93                                  & 2.43                                  & 1.74                                  & 0.89                                  & 9.42                                  & 1.73                                  & 4.83                                  \\ 
\multicolumn{1}{c|}{\multirow{-3}{*}{\textbf{OOD}}}  & w Single-Stream Encoder              & 24.91                                  & 2.93                                  & 6.88                                  & 27.15                                  & 3.05                                  & 7.02                                  & 6.48                                  & 1.50                                  & 2.76                                  & 2.58                                  & 1.93                                  & 1.20                                  & 9.06                                  & 1.78                                  & 3.01                                  \\ \toprule
\end{tabular}}
\end{table*}

%% file: tab/ablation_recon.tex
\begin{table*}[th]
\caption{
    \textbf{Ablations on the monologic-anchored motion factorization learning scheme.}
    Reconstruction performance under different loss configurations, validating the necessity of each component for effective semantics-interaction disentanglement and demonstrating their complementary contributions to stable representation learning.
}
\label{ablation_recon}
\resizebox{\textwidth}{!}{\begin{tabular}{@{}cc|ccc|ccc|ccc|ccc|ccc@{}}
\bottomrule
\multicolumn{2}{c|}{}                                                                      & \multicolumn{3}{c|}{\textbf{FD $\downarrow$}}                                                                         & \multicolumn{3}{c|}{\textbf{P-FD $\downarrow$}}                                                                       & \multicolumn{3}{c|}{\textbf{MSE $\downarrow$}}                                                                        & \multicolumn{3}{c|}{\textbf{SID $\uparrow$}}                                                                                   & \multicolumn{3}{c}{\textbf{rPCC $\downarrow$}}                                                               \\ \cline{3-17} 
\multicolumn{2}{c|}{\multirow{-2}{*}{\textbf{Methods}}}                                    & \textbf{exp}                          & \textbf{jaw}                          & \textbf{pose}                         & \textbf{exp}                          & \textbf{jaw}                          & \textbf{pose}                         & \textbf{exp}                          & \textbf{jaw}                          & \textbf{pose}                         & \textbf{exp}                          & \textbf{jaw}                          & \textbf{pose}                         & \textbf{exp}                          & \textbf{jaw}                          & \textbf{pose}                         \\ \hline
\multicolumn{1}{c|}{}                                & \cellcolor[HTML]{F2F2F2}Ours(Recon) & \cellcolor[HTML]{F2F2F2}\textbf{7.44} & \cellcolor[HTML]{F2F2F2}\textbf{0.40} & \cellcolor[HTML]{F2F2F2}\textbf{2.51} & \cellcolor[HTML]{F2F2F2}\textbf{8.16} & \cellcolor[HTML]{F2F2F2}\textbf{0.45} & \cellcolor[HTML]{F2F2F2}\textbf{2.48} & \cellcolor[HTML]{F2F2F2}\textbf{2.80} & \cellcolor[HTML]{F2F2F2}\textbf{0.56} & \cellcolor[HTML]{F2F2F2}\textbf{0.86} & \cellcolor[HTML]{F2F2F2}\textbf{3.66} & \cellcolor[HTML]{F2F2F2}\textbf{2.75} & \cellcolor[HTML]{F2F2F2}\textbf{2.21} & \cellcolor[HTML]{F2F2F2}\textbf{3.31} & \cellcolor[HTML]{F2F2F2}\textbf{0.57} & \cellcolor[HTML]{F2F2F2}\textbf{1.84} \\
\multicolumn{1}{c|}{}                                & w/o $\mathcal{L}_{rec}$             & 8.92                                  & 0.58                                  & 2.97                                  & 9.61                                  & 0.62                                  & 2.95                                  & 3.64                                  & 0.72                                  & 1.09                                  & 3.21                                  & 2.40                                  & 1.88                                  & 4.12                                  & 0.73                                  & 2.21                                  \\
\multicolumn{1}{c|}{}                                & w/o $\mathcal{L}_\text{swap}$       & 8.37                                  & 0.49                                  & 2.83                                  & 9.25                                  & 0.54                                  & 2.79                                  & 3.18                                  & 0.63                                  & 0.98                                  & 3.48                                  & 2.61                                  & 2.02                                  & 3.78                                  & 0.69                                  & 2.05                                  \\
\multicolumn{1}{c|}{}                                & w/o $\mathcal{L}_\text{sem-cyc}$   & 8.17                                  & 0.47                                  & 2.72                                  & 8.93                                  & 0.51                                  & 2.70                                  & 3.09                                  & 0.60                                  & 0.92                                  & 3.42                                  & 2.64                                  & 2.05                                  & 3.59                                  & 0.65                                  & 1.98                                  \\
\multicolumn{1}{c|}{}                                & w/o $\mathcal{L}_\text{int-cyc}$    & 8.26                                  & 0.47                                  & 2.79                                  & 9.04                                  & 0.53                                  & 2.75                                  & 3.12                                  & 0.62                                  & 0.95                                  & 3.30                                  & 2.56                                  & 1.99                                  & 3.60                                  & 0.70                                  & 2.07                                  \\ 
\multicolumn{1}{c|}{\multirow{-6}{*}{\textbf{Test}}} & w/o $\mathcal{L}_\text{KL}$         & 7.68                                  & 0.41                                  & 2.56                                  & 8.36                                  & 0.46                                  & 2.53                                  & 2.88                                  & 0.58                                  & 0.88                                  & 3.74                                  & 2.81                                  & 2.29                                  & 3.44                                  & 0.60                                  & 1.91                                  \\ \hline \hline
\multicolumn{1}{c|}{}                                & \cellcolor[HTML]{F2F2F2}Ours(Recon) & \cellcolor[HTML]{F2F2F2}\textbf{8.01} & \cellcolor[HTML]{F2F2F2}\textbf{0.46} & \cellcolor[HTML]{F2F2F2}\textbf{3.07} & \cellcolor[HTML]{F2F2F2}\textbf{8.71} & \cellcolor[HTML]{F2F2F2}\textbf{0.51} & \cellcolor[HTML]{F2F2F2}\textbf{3.12} & \cellcolor[HTML]{F2F2F2}\textbf{3.00} & \cellcolor[HTML]{F2F2F2}\textbf{0.60} & \cellcolor[HTML]{F2F2F2}\textbf{1.11} & \cellcolor[HTML]{F2F2F2}\textbf{3.68} & \cellcolor[HTML]{F2F2F2}\textbf{2.74} & \cellcolor[HTML]{F2F2F2}\textbf{2.10} & \cellcolor[HTML]{F2F2F2}\textbf{3.75} & \cellcolor[HTML]{F2F2F2}\textbf{0.63} & \cellcolor[HTML]{F2F2F2}\textbf{2.04} \\
\multicolumn{1}{c|}{}                                & w/o $\mathcal{L}_{rec}$             & 9.84                                  & 0.61                                  & 3.64                                  & 10.66                                 & 0.64                                  & 3.58                                  & 3.92                                  & 0.79                                  & 1.34                                  & 3.18                                  & 2.28                                  & 1.79                                  & 4.63                                  & 0.81                                  & 2.47                                  \\
\multicolumn{1}{c|}{}                                & w/o $\mathcal{L}_\text{swap}$       & 9.22                                  & 0.55                                  & 3.41                                  & 10.10                                 & 0.60                                  & 3.35                                  & 3.58                                  & 0.72                                  & 1.22                                  & 3.39                                  & 2.42                                  & 1.86                                  & 4.24                                  & 0.75                                  & 2.32                                  \\
\multicolumn{1}{c|}{}                                & w/o $\mathcal{L}_\text{sem-cyc}$   & 8.94                                  & 0.52                                  & 3.29                                  & 9.78                                  & 0.57                                  & 3.22                                  & 3.41                                  & 0.68                                  & 1.15                                  & 3.44                                  & 2.45                                  & 1.93                                  & 4.02                                  & 0.72                                  & 2.24                                  \\
\multicolumn{1}{c|}{}                                & w/o $\mathcal{L}_\text{int-cyc}$    & 9.05                                  & 0.53                                  & 3.35                                  & 9.92                                  & 0.59                                  & 3.28                                  & 3.47                                  & 0.70                                  & 1.18                                  & 3.29                                  & 2.38                                  & 1.90                                  & 4.10                                  & 0.76                                  & 2.30                                  \\ 
\multicolumn{1}{c|}{\multirow{-6}{*}{\textbf{OOD}}}  & w/o $\mathcal{L}_\text{KL}$         & 8.24                                  & 0.47                                  & 3.13                                  & 8.94                                  & 0.52                                  & 3.18                                  & 3.09                                  & 0.62                                  & 1.14                                  & 3.78                                  & 2.85                                  & 2.23                                  & 3.88                                  & 0.67                                  & 2.12                                  \\ \toprule
\end{tabular}}
\end{table*}

%% file: sec/conclusion.tex
\section{Limitations and Future Work}
A current limitation is that the framework focuses on head motion and models interaction primarily through audio signals. Future work will explore extending the approach to richer conversational behaviors, such as full-body gestures, gaze dynamics, and multimodal conversational cues, toward more comprehensive and socially aware interactive avatars. 
\section{Conclusion}
We present Learn2Chat, a unified framework for audio-driven dyadic head motion generation that models conversational behavior as interaction modulation over pretrained monologic motion priors. By explicitly separating speech-driven motion dynamics from partner-conditioned interaction signals, Learn2Chat alleviates the signal conflation inherent in holistic dyadic learning and produces temporally stable, synchronized conversational motions. The framework distills clean interaction representations from dyadic motion via a monologic-anchored factorization scheme and predicts interaction latents directly from paired audio through cross-attentive coupling. Experiments on the DualTalk benchmark demonstrate consistent improvements over prior methods in motion fidelity, diversity, and inter-speaker coordination.

%% file: sec/biography.tex
\begin{IEEEbiography}[{\includegraphics[width=1in,height=1.25in,clip,keepaspectratio]{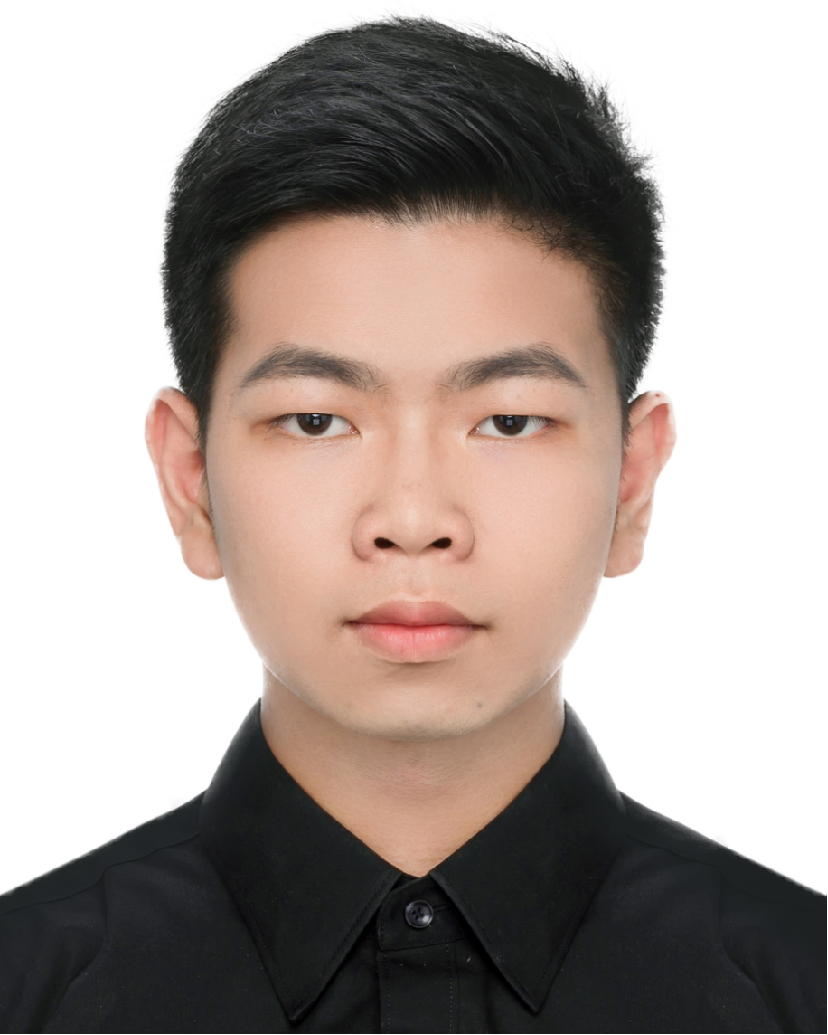}}]{Zikai Huang} is currently pursuing a Ph.D. degree in the School of Computer Science and Engineering, South China University of Technology. His research interests include computer vision, computer graphics and multimodal learning.
\end{IEEEbiography}

\begin{IEEEbiography}[{\includegraphics[width=1in,height=1.25in,clip,keepaspectratio]{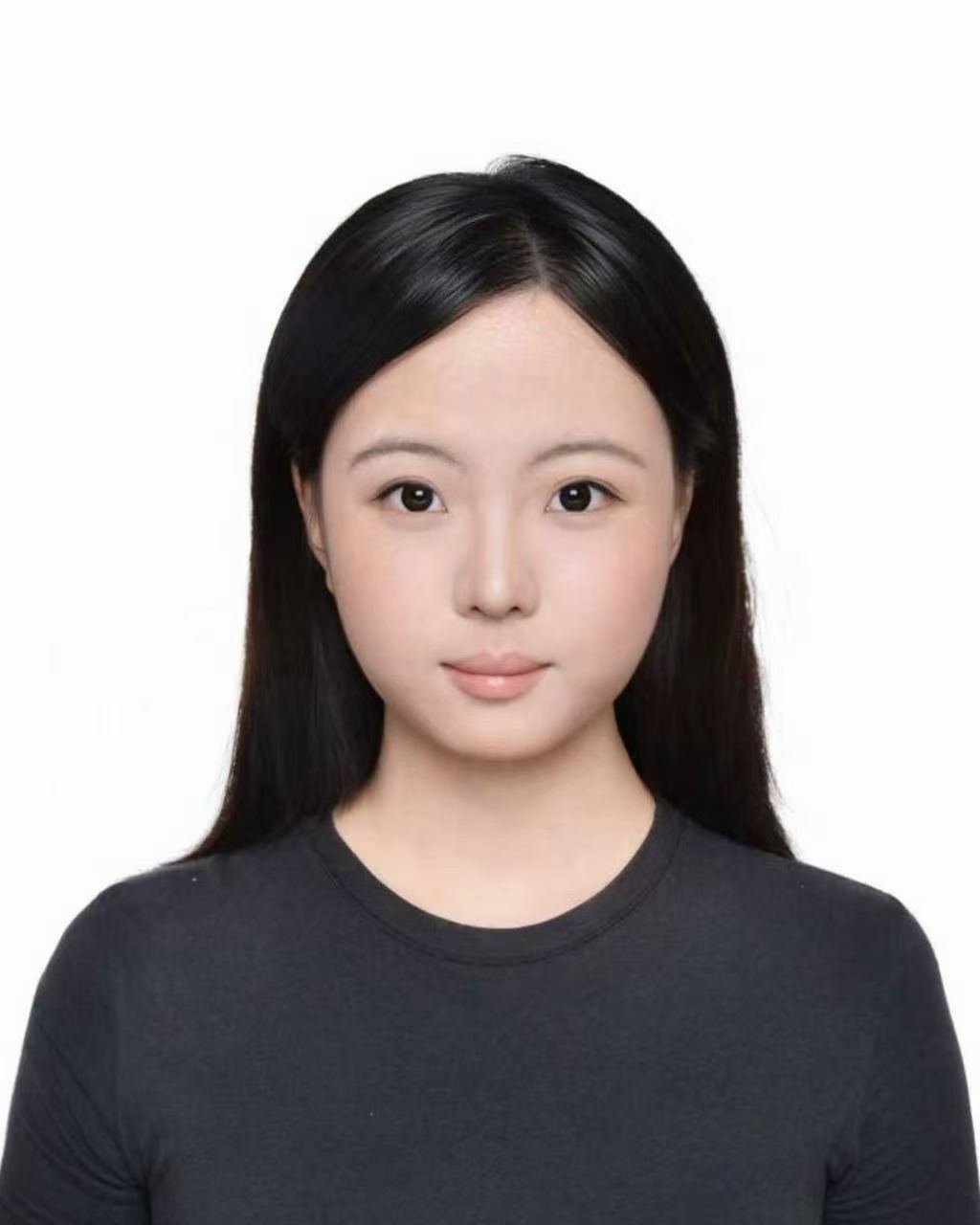}}]{Siyue Chen} is currently an undergraduate student in the School of Design, South China University of Technology. Her research interests include 3D vision, mechanistic interpretability, and language models.
\end{IEEEbiography}

\begin{IEEEbiography}[{\includegraphics[width=1in,height=1.25in,clip,keepaspectratio]{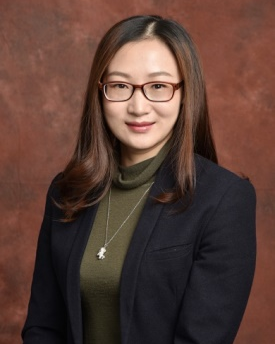}}]{Xuemiao Xu} received the BS and MS degrees in computer science and engineering from South China University of Technology, in 2002 and 2005, respectively, and the PhD degree in computer science and engineering from The Chinese University of Hong Kong in 2009. She is currently a professor with the School of Computer Science and Engineering, South China University of Technology. Her research interests include object detection, tracking, recognition, and image, video understanding and synthesis, particularly their applications in the intelligent transportation.
\end{IEEEbiography}

\begin{IEEEbiography}[{\includegraphics[width=1in,height=1.25in,clip,keepaspectratio]{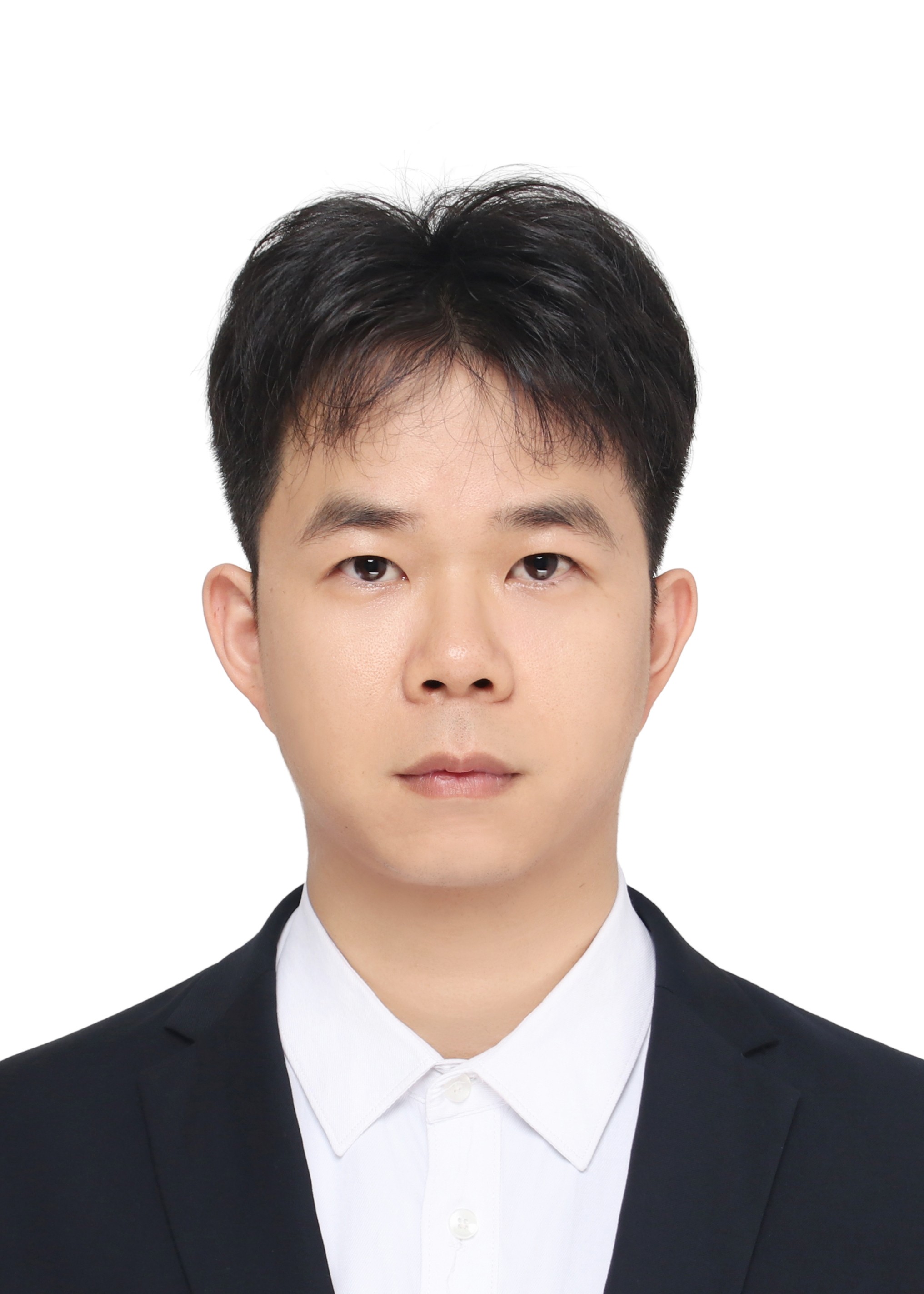}}]{Haoxin Yang} received his Ph.D. degree from the School of Computer Science \& Engineering, South China University of Technology. He obtained his B.Sc. and M.Sc. degrees from South China Agricultural University and Shenzhen University in 2019 and 2022, respectively. \end{IEEEbiography}

\begin{IEEEbiography}[{\includegraphics[width=1in,height=1.25in,clip,keepaspectratio]{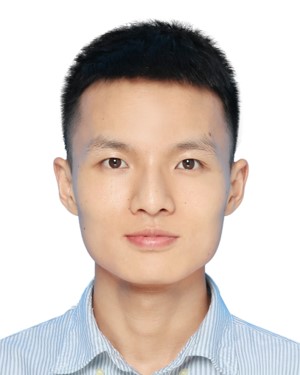}}]{Cheng Xu} received his Ph.D. degree in computer science and technology from South China University of Technology, China, in 2023. He is currently a Research Scientist at Singapore Management University. Prior to this, he was a Post-Doctoral Fellow at The Hong Kong Polytechnic University from 2023 to 2026. His research interests primarily include human-centric visual representation, understanding, and generation.
\end{IEEEbiography}

\begin{IEEEbiography}[{\includegraphics[width=1in,height=1.25in,clip,keepaspectratio]{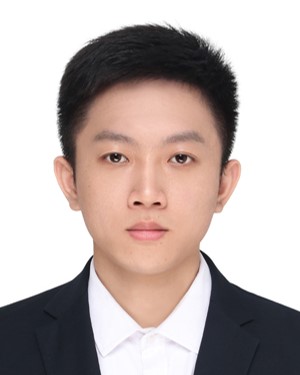}}]{Yihong Lin} received the B.Sc. and M.Sc. degrees from South China University of Technology in 2022 and 2025, respectively. He is currently pursuing the Ph.D. degree with the School of Computer Science and Engineering at the same university. His research interests include 3D reconstruction and generative modeling.
\end{IEEEbiography}

\begin{IEEEbiography}[{\includegraphics[width=1in,height=1.25in,clip,keepaspectratio]{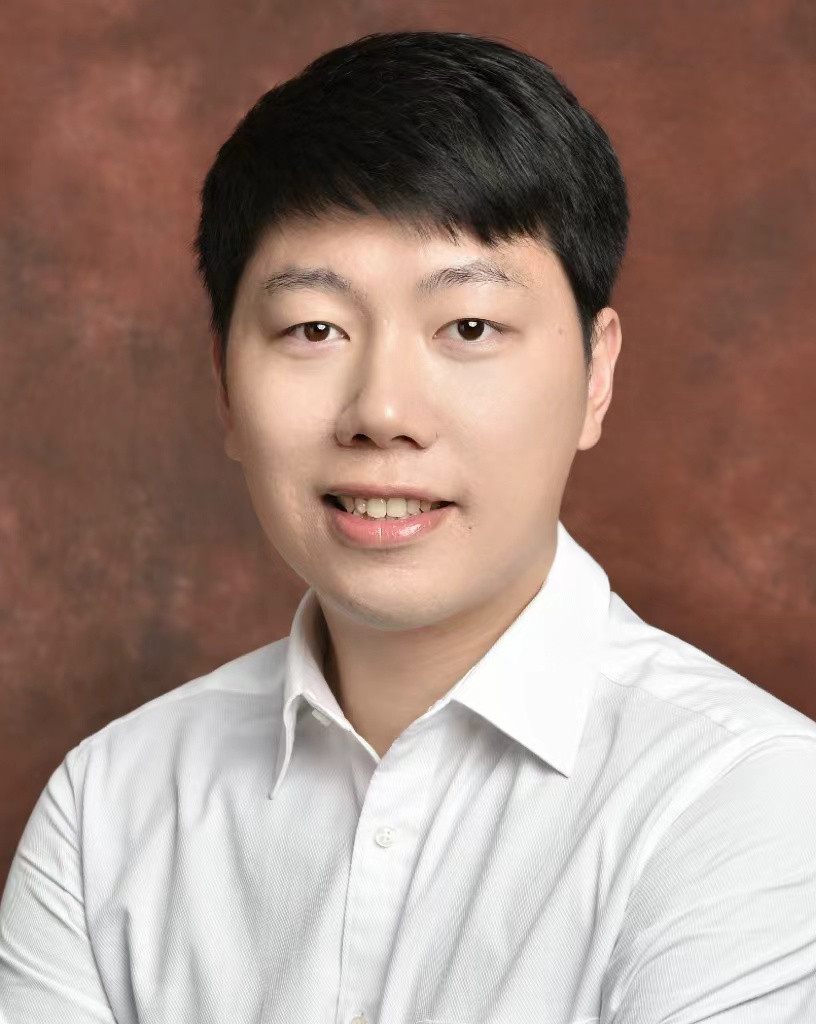}}]{Shengfeng He} (Senior Member, IEEE) is an associate professor in the School of Computing and Information Systems at Singapore Management University. He earned his B.Sc. and M.Sc. from Macau University of Science and Technology (2009, 2011) and a Ph.D. from City University of Hong Kong (2015). His research focuses on computer vision and generative models. He has received awards including the Google Research Award (2024), PerCom 2024 Best Paper Award, and the Lee Kong Chian Fellowship (2024, 2026). He is a senior IEEE member and a distinguished CCF member. He serves as the Associate Editor-in-Chief of Neurocomputing, lead guest editor for IJCV, and associate editor for IEEE TPAMI. He is a senior area chair for NeurIPS, an area chair for CVPR, NeurIPS, ICLR, ICML, AAAI, IJCAI, and the Conference Chair of Pacific Graphics 2026.
\end{IEEEbiography} 

%% file: suppl/tab/listener.tex
\begin{table*}[t!]
    \caption{
        \textbf{Evaluation on listener motion generation.}
        Learn2Chat consistently improves listener motion fidelity and diversity over monologic baselines on the Test/OOD set, demonstrating responsive and context-aware interaction behaviors.
    }
    \resizebox{1.0\linewidth}{!}{
    \renewcommand{\arraystretch}{1.2}
    \setlength{\tabcolsep}{1pt}
    \begin{tabular}{@{}c|ccc|ccc|ccc|ccc|ccc@{}}
    \bottomrule
    \multirow{3}{*}{\textbf{Methods}}       & \multicolumn{3}{c|}{\textbf{FD $\downarrow$}}                                                                                                          & \multicolumn{3}{c|}{\textbf{P-FD $\downarrow$}}                                                                                                        & \multicolumn{3}{c|}{\textbf{MSE $\downarrow$}}                                                                                                                                                       & \multicolumn{3}{c|}{\textbf{SID $\uparrow$}}    & \multicolumn{3}{c}{\textbf{rPCC $\downarrow$}}                                                                                                                                                       \\ \cline{2-16}
                                            & exp               & \begin{tabular}[c]{@{}c@{}}jaw\\ $\times 10^{-3}$ \end{tabular} & \begin{tabular}[c]{@{}c@{}}pose\\ $\times 10^{-2}$ \end{tabular} & exp               & \begin{tabular}[c]{@{}c@{}}jaw\\ $\times 10^{-3}$ \end{tabular} & \begin{tabular}[c]{@{}c@{}}pose\\ $\times 10^{-2}$ \end{tabular} & \begin{tabular}[c]{@{}c@{}}exp\\ $\times 10^{-1}$ \end{tabular} & \begin{tabular}[c]{@{}c@{}}jaw\\ $\times 10^{-3}$ \end{tabular} & \begin{tabular}[c]{@{}c@{}}pose\\ $\times 10^{-2}$ \end{tabular} & exp              & jaw              & pose      & \begin{tabular}[c]{@{}c@{}}exp\\ $\times 10^{-2}$ \end{tabular} & \begin{tabular}[c]{@{}c@{}}jaw\\ $\times 10^{-1}$ \end{tabular} & \begin{tabular}[c]{@{}c@{}}pose\\ $\times 10^{-1}$ \end{tabular} \\ \hline
    DEEPTalk~\cite{kim2025deeptalk}         & 30.98/31.54       & 5.31/5.26                                                       & 3.00/2.85                                                        & 31.02/31.82       & 5.31/5.27                                                       & 3.24/3.07                                                        & 6.72/6.81                                                       & 1.76/1.75                                                       & 2.05/2.04                                                        & 1.42/1.46        & 1.44/1.32        & 1.67/1.68 & 15.88/15.35                                                     & 2.48/2.16                                                       & 2.09/1.85                                                        \\
    UniTalker~\cite{fan2024unitalker}       & 30.79/31.08       & 5.30/5.28                                                       & 2.91/2.79                                                        & 30.79/31.07       & 5.30/5.26                                                       & 3.14/3.01                                                        & 6.74/6.82                                                       & 1.77/1.75                                                       & 1.85/1.86                                                        & 1.43/1.51        & 1.39/1.38        & 1.72/1.74 & 15.75/15.18                                                     & 2.49/2.12                                                       & 2.58/2.12                                                        \\ \hline
    Ours(DEEPTalk~\cite{kim2025deeptalk})   & 15.50/22.85       & 1.80/2.62                                                       & 2.13/2.01                                                        & 16.61/23.98       & 1.91/2.72                                                       & 2.36/2.23                                                        & 4.53/6.14                                                       & 1.11/1.41                                                       & 1.57/1.59                                                        & 3.02/2.70        & 2.22/2.04        & 1.87/1.85 & 6.89/7.70                                                       & 1.38/1.49                                                       & 1.62/1.41                                                        \\
    Ours(UniTalker~\cite{fan2024unitalker}) & 15.08/22.97       & 1.80/2.67                                                       & 2.43/2.34                                                        & 16.23/24.14       & 1.91/2.78                                                       & 2.67/2.56                                                        & 4.55/6.29                                                       & 1.12/1.44                                                       & 1.71/1.73                                                        & 3.05/2.73        & 2.23/2.01        & 1.75/1.70 & 6.62/7.46                                                       & 1.39/1.51                                                       & 1.57/1.38                                                        \\ 
    \toprule
    \end{tabular}}
    \label{listener}
\end{table*} 

%% file: suppl/tab/generalization.tex
\begin{table*}[t!]
    \caption{
        \textbf{Cross-backbone generalization of Learn2Chat.}
        Ours([A]@[B]) denotes training with backbone A and inference with backbone B on the Test/OOD set.
        The stable performance across different backbone combinations demonstrates that the learned interaction modulation is transferable and backbone-agnostic.
    }
    \resizebox{1.0\linewidth}{!}{
    \renewcommand{\arraystretch}{1.2}
    \setlength{\tabcolsep}{1pt}
    \begin{tabular}{@{}c|ccc|ccc|ccc|ccc|ccc@{}}
    \bottomrule
    \multirow{3}{*}{\textbf{Methods}}                                        & \multicolumn{3}{c|}{\textbf{FD $\downarrow$}}                                                                                                          & \multicolumn{3}{c|}{\textbf{P-FD $\downarrow$}}                                                                                                        & \multicolumn{3}{c|}{\textbf{MSE $\downarrow$}}                                                                                                                                                       & \multicolumn{3}{c|}{\textbf{SID $\uparrow$}}    & \multicolumn{3}{c}{\textbf{rPCC $\downarrow$}}                                                                                                                                                       \\ \cline{2-16}
                                        & exp                                & \begin{tabular}[c]{@{}c@{}}jaw\\ $\times 10^{-3}$ \end{tabular} & \begin{tabular}[c]{@{}c@{}}pose\\ $\times 10^{-2}$ \end{tabular} & exp               & \begin{tabular}[c]{@{}c@{}}jaw\\ $\times 10^{-3}$ \end{tabular} & \begin{tabular}[c]{@{}c@{}}pose\\ $\times 10^{-2}$ \end{tabular} & \begin{tabular}[c]{@{}c@{}}exp\\ $\times 10^{-1}$ \end{tabular} & \begin{tabular}[c]{@{}c@{}}jaw\\ $\times 10^{-3}$ \end{tabular} & \begin{tabular}[c]{@{}c@{}}pose\\ $\times 10^{-2}$ \end{tabular} & exp              & jaw              & pose      & \begin{tabular}[c]{@{}c@{}}exp\\ $\times 10^{-2}$ \end{tabular} & \begin{tabular}[c]{@{}c@{}}jaw\\ $\times 10^{-1}$ \end{tabular} & \begin{tabular}[c]{@{}c@{}}pose\\ $\times 10^{-1}$ \end{tabular} \\ \hline
    Ours(DEEPTalk~\cite{kim2025deeptalk}@UniTalker~\cite{fan2024unitalker})  & 15.05/22.66       & 2.23/3.03                                                       & 6.01/6.12                                                        & 16.09/23.76       & 2.33/3.13                                                       & 5.98/6.44                                                        & 4.63/6.32                                                       & 1.31/1.60                                                       & 2.34/2.64                                                        & 3.07/2.75        & 2.09/1.92        & 1.80/1.75 & 6.87/7.63                                                       & 1.46/1.52                                                       & 2.70/2.52                                                        \\
Ours(UniTalker~\cite{fan2024unitalker}@DEEPTalk~\cite{kim2025deeptalk})      & 15.78/23.00       & 1.86/2.54                                                       & 5.46/5.65                                                        & 16.80/24.07       & 1.96/2.64                                                       & 5.82/5.82                                                        & 4.68/6.27                                                       & 1.18/1.42                                                       & 2.27/2.53                                                        & 3.07/2.76        & 2.24/2.05        & 1.39/1.36 & 6.96/7.91                                                       & 1.39/1.50                                                       & 2.59/2.49                                                        \\
    \toprule
    \end{tabular}}
    \label{generation}
\end{table*} 

%% file: suppl/intro.tex
In this supplementary material, we provide additional implementation details of Learn2Chat in~\refsec{implementation}, including motion and audio representations as well as training configurations. We further  present additional analyses of role-specific interaction behaviors in~\refsec{listener} and evaluate the cross-backbone generalization capability of Learn2Chat in~\refsec{generalization}, demonstrating the effectiveness of the learned interaction modulation across different pretrained monologic motion models.

%% file: suppl/implementation_details.tex
\section{Implementation Details}\label{sec:implementation}
\subsection{Motion and Audio Representation}
Following prior works~\cite{ng2022learning,tran2024dim,peng2025dualtalk,kim2025deeptalk}, we represent 3D head motion using the FLAME model~\cite{li2017learning}, which provides a widely adopted parametric representation for facial and head dynamics. 
To better model the dynamics of neck pose movements, we predict the relative rotation of the neck pose with respect to the first frame of the sequence, rather than the absolute rotation.
When inferencing the head motion, we can reconstruct the absolute neck pose by accumulating the predicted relative rotations over time base on the initial neck pose.
This formulation enables consistent modeling of head motion trajectories across different sequences.
For audio representation, we extract speech features using the pretrained Wav2Vec2.0 model~\cite{baevski2020wav2vec}.
These features capture high-level speech characteristics and have been demonstrated to be effective for audio-driven motion generation in previous studies~\cite{fan2024unitalker,peng2025dualtalk,peng2024synctalk,sun2025vividtalk,he2024emotalk3d,wang2025emotivetalk}.

\subsection{Training Details}
All experiments are implemented in PyTorch 2.4.1 and trained on 4 NVIDIA 4090D GPUs. 
We adopt the AdamW optimizer~\cite{loshchilov2017decoupled} with a learning rate of 1e-4 and a batch size of 64.
During training, motion sequences are sampled using a fixed non-overlap temporal window of 200 frames with a frame rate of 25 fps.
The training process follows a two-stage strategy.
We first pretrain the Semantics Encoder and Semantics-Interaction Decoder to establish a robust disentangled representation that effectively captures the underlying semantics of speech and interaction dynamics.
This pretraining phase is crucial for ensuring that the subsequent interaction modeling can effectively leverage the learned representations and is trained for 3000 epochs.
After the pretraining phase, we fix the parameters of the Semantics Encoder and Semantics-Interaction Decoder to preserve the learned disentangled representations.
We then proceed to train the Audio-to-Interaction Encoder for an additional 1000 epochs.

To balance the training of different loss components, we adopt an automatic weighted mechanism~\cite{liebel2018auxiliary,chen2025clip}.
The adaptive loss weight for each loss term is defined as follows:
\begin{equation}
\lambda_{i} = \frac{1}{2\omega_{i}^2} \mathcal{L}_{i} + \log (1 + \omega_{i}),
\end{equation}
where $\omega_{i}$ denotes a learnable weight parameter for each loss term $\mathcal{L}_{i}$.
The adaptive loss weight strategy allows the model to automatically balance the contributions of different loss components during training, leading to improved convergence and better overall performance.

%% file: suppl/listener.tex
\section{Analysis of Listener Behavior} \label{sec:listener}
To further analyze the granularity of learned interaction behaviors, we utilize separate audio tracks to infer speaker and listener roles and report role-specific evaluation metrics.
As shown in~\reftab{listener}, our method consistently improves performance on listener-role motion generation compared with monologic baselines.
These improvements indicate that the model is able to generate responsive and context-aware listener motions rather than defaulting to static or purely speech-driven behaviors during non-speaking intervals. In particular, the gains in both motion fidelity and diversity suggest that listener-side motions are dynamically modulated by conversational context, reflecting more adaptive interaction-aware generation.

%% file: suppl/generalization.tex
\section{Generalization Across Monologic Backbones} \label{sec:generalization}
We further evaluate the generalization capability of Learn2Chat across different monologic motion predictors.
As shown in~\reftab{generation}, we train Learn2Chat with one pretrained monologic generator and replace it with another at inference time.
The results remain stable across different backbone combinations, demonstrating that the learned interaction modulation is not tied to a specific monologic model.
This suggests that Learn2Chat operates on a transferable speech-semantic motion space and can effectively adapt interaction modeling across different pretrained motion priors without retraining.